\documentclass[12pt, leqno]{article}
\usepackage{amsmath,setspace,multirow,lineno}
\usepackage[font=small,labelfont=bf]{caption}
\usepackage[top=1.3in, bottom=1.3in, left=1.1in, right=1.1in]{geometry}

\usepackage[round]{natbib}
\usepackage{color,soul}

\DeclareCaptionStyle{italic}[justification=centering]{labelfont={bf},textfont={it},labelsep=colon}
\usepackage{graphicx,psfrag,epsf,textcomp,epstopdf, amsthm, paralist}
\usepackage{enumerate, dsfont, alltt, verbatim}
\usepackage{natbib}
\usepackage{url,xcolor}
\usepackage{booktabs, subfig, bm, paralist,mathpazo,tikz,todonotes,longtable,microtype}
\usepackage[linesnumbered,ruled,vlined]{algorithm2e}

\usepackage[pdftex,colorlinks=true]{hyperref}
\definecolor{darkblue}{rgb}{0,0,.6}
\hypersetup{citecolor=darkblue,linkcolor=darkblue,urlcolor=darkblue}
\definecolor{DarkRed}{rgb}{.7,0,.4}

\newcommand{\argmax}{\operatornamewithlimits{argmax}}

\newcommand{\blind}{0}

\newcommand{\X}{\mathcal{X}}

\newcommand{\Rlogo}{\protect\includegraphics[height=1.8ex,keepaspectratio]{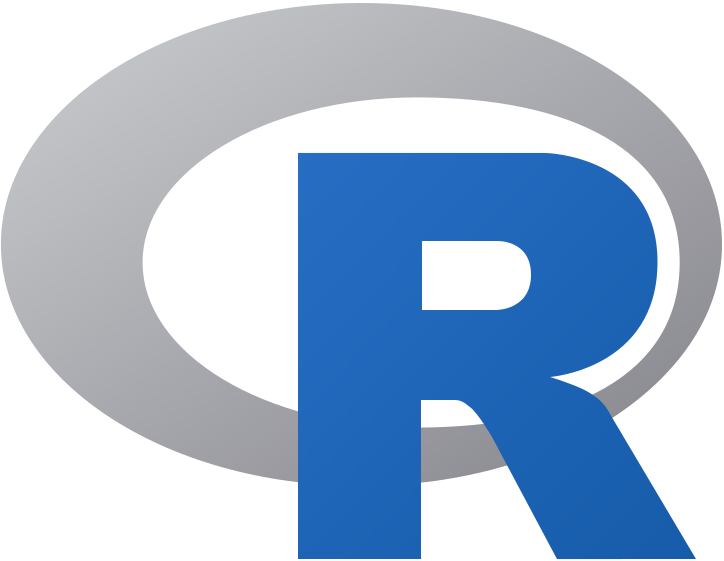}}

\graphicspath{{plots/}}
\DeclareMathOperator*{\argmin}{\arg\!\min}

\newsavebox\CBox

 \newtheorem{@definition}{\sc Definition}[section]

 \newtheorem{lemma}{\sc Lemma}[section]
 \newtheorem{remark}{\sc Remark}[section]

  \renewcommand\X{\mathcal{X}}

\date{}
\begin{document}

\def\spacingset#1{\renewcommand{\baselinestretch}{#1}\small\normalsize} \spacingset{1}

\if0\blind
{
\title{\bf Robust Functional Logistic Regression}}
\author{
\normalsize Berkay Akturk
\\
\normalsize Graduate School of Natural and Applied Sciences \\
\normalsize Marmara University \\
\\
\normalsize Ufuk Beyaztas\footnote{Corresponding address: Department of Statistics, Marmara University, 34722, Kadikoy-Istanbul, Turkey; Email: ufuk.beyaztas@marmara.edu.tr} 
\\
\normalsize Department of Statistics \\
\normalsize Marmara University \\
\\
\normalsize Han Lin Shang 
\\
\normalsize Department of Actuarial Studies and Business Analytics \\
\normalsize Macquarie University
}
\maketitle
\fi

\if1\blind
{
\title{\bf Robust Functional Logistic Regression}
} \fi

\maketitle

\begin{abstract}
Functional logistic regression is a popular model to capture a linear relationship between binary response and functional predictor variables. However, many methods used for parameter estimation in functional logistic regression are sensitive to outliers, which may lead to inaccurate parameter estimates and inferior classification accuracy. We propose a robust estimation procedure for functional logistic regression, in which the observations of the functional predictor are projected onto a set of finite-dimensional subspaces via robust functional principal component analysis. This dimension-reduction step reduces the outlying effects in the functional predictor. The logistic regression coefficient is estimated using an M-type estimator based on binary response and robust principal component scores. In doing so, we provide robust estimates by minimizing the effects of outliers in the binary response and functional predictor variables. Via a series of Monte-Carlo simulations and using hand radiograph data, we examine the parameter estimation and classification accuracy for the response variable. We find that the robust procedure outperforms some existing robust and non-robust methods when outliers are present, while producing competitive results when outliers are absent. In addition, the proposed method is computationally more efficient than some existing robust alternatives.
\end{abstract}

\noindent \textit{Keywords}: Bianco and Yohai estimator, Functional data, Functional principal component analysis, Logistic regression. 

\newpage
\spacingset{1.55} 

\section{Introduction} \label{sec:intro}

Functional data analysis techniques enable modeling data observed over a continuum, such as time, space, depth, and wavelength. These techniques can be used for a wide range of purposes, including estimation, classification, and regression of the functional form of the data. We study the estimation of the parameters in a functional logistic regression (FlogitR), which involves a binary response variable and a predictor variable composed of random curves.

Several existing studies for parameter estimation and classification of the response variable in the FlogitR approximate the functional predictor variable and regression coefficient function to a finite-dimensional space using a set of pre-determined bases, such as Gaussian, Fourier and B-spline. For example, \cite{ratcliffe2002functional} propose an estimation method for FlogitR using Fourier expansions of the functional predictor variable and regression coefficient function. \cite{araki2009functional} perform parameter estimation of FlogitR using the Gaussian basis expansion method.

Functional principal component analysis (FPCA) and functional partial least squares (FPLS) are data-driven dimension-reduction methods that have been extended to the estimation of the FlogitR to better capture the variability in the data and estimate model parameters more effectively. FPCA projects the functional predictors onto a set of finite-dimensional subspaces, obtaining a set of orthonormal principal components and their associated scores. This method considers maximizing the covariance between the predictor variables when obtaining the principal components. Subsequently, the regression coefficient function in FlogitR is approximated using a logistic regression model constructed between the principal component scores and the binary response variable. Like FPCA, FPLS projects the functional predictors onto a finite-dimensional subspace. Unlike the FPCA method, the FPLS method extracts latent components that maximize covariance between the response and predictor variables. For parameter estimation, classification, and model selection, various methods based on FPCA have been proposed in FlogitR by \cite{escabias2004principal}, \cite{escabias2005modeling}, \cite{leng2006classification}, \cite{aguilera2008discussion}, \cite{wei2014functional}, and, \cite{mousavi2017multinomial}, and their findings have been compared with those obtained using classical logistic regression models. \cite{preda2007pls} and \cite{escabias2007functional} propose FPLS-based methods for estimating the parameters in FlogitR. Additionally, \cite{mousavi2018functional} compare the performance of estimators based on FPCA, FPLS, and LASSO methods for FlogitR.

The findings from these studies demonstrate that the FPCA and FPLS methods are more effective than several pre-determined basis expansion approaches in predicting the response variable from FlogitR. This is because the pre-determined basis expansion may require a large number of bases to project infinite-dimensional functional objects onto a finite-dimensional space \citep{matsui2009regularized, beyaztas2020function}, leading to model overfitting of the model and decreased classification accuracy. 

Although the FPCA and FPLS-based estimation methods produce improved results over the pre-determined basis expansion when data follow a smooth data generation process, these methods are sensitive to outliers in the dataset. In the presence of outliers, these methods may produce biased parameter estimates and decreased classification accuracy. To our knowledge, only two methods have been proposed to address this issue. \cite{denhere2016robust} integrate the minimum covariance determinant approach into the FPCA method to obtain principal component scores. They then obtained the estimates of parameters constructed using binary response and principal component scores by the maximum likelihood estimator, which is used to approximate the regression coefficient function in the FlogitR. However, this method is only partially robust because the maximum likelihood estimator is not robust to outliers. In contrast, \cite{mutis2022robust} propose a novel robust FPLS-based method (RFPLS) to estimate FlogitR in the presence of outliers. In this method, FPLS components are first obtained based on the weighted likelihood principle. Then, the parameters of the logistic regression model constructed using binary response and FPLS scores are estimated by the weighted likelihood-based generalized linear model. Through numerical analysis, \cite{mutis2022robust} showed that the RFPLS method produces robust estimates against outliers in predictor and response variables. However, a disadvantage of this method is that the iteration-based estimation method used in the weighted likelihood method significantly increases computing time. For example, our results, discussed in Section~\ref{sec:nr} in more detail, show that the RFPLS method requires 40 to 1200 times more computing time than non-robust methods.

The proposed method enables efficient and fast model estimation in FlogitR in the presence of outliers. In the proposed method, the functional predictors are projected onto a finite-dimensional subspace using the robust functional principal component analysis (RFPCA) method proposed by \cite{bali2011rfpc}. The RFPCA method resiliently projects functional observations into a finite-dimensional space using a robust scale estimator and the robust projection pursuit approach proposed by \cite{croux96}. Subsequently, the parameter of the logistic regression model constructed using the RFPCA scores and the binary response variable is estimated by a procedure proposed by \cite{croux2003implementing}. This procedure employs the weighted \citeauthor{bianco1996robust}'s \citeyearpar{bianco1996robust} estimator, which is robust to outliers. Therefore, our proposed method produces robust predictions even when outliers exist in both response and predictor variables. 

Unlike the weighted likelihood method, which uses resampling techniques when optimizing the parameter estimates, our proposed method is computationally efficient because it neither requires the use of resampling methods nor many iterations in the estimation of the regression coefficient. Our numerical analyses show that, compared with non-robust methods, the proposed method produces competitive results when no outliers are present in the data. On the other hand, when outliers are present in the data, the proposed method provides improved parameter estimates and classification accuracy. In addition, our results show that the proposed method produces competitive or improved parameter estimation and classification accuracy compared to the RFPLS method, with 3 to 151 times less computation time. Under some regularity conditions, we show the asymptotic consistency of the proposed estimator (please see the Appendix). In addition, the proposed estimator's influence function is derived by leveraging the influence functions of RFPCA and weighted \citeauthor{bianco1996robust}'s \citeyearpar{bianco1996robust} estimators, which shows that only good leverage points may affect the proposed estimator. In addition, the asymptotic distribution of the proposed estimator is derived through the utilization of the Bahadur expansion.

The remainder of this paper is organized as follows. FlogitR and the general principle used to estimate the model parameters are summarized in Section~\ref{sec:FlogitR}. The proposed method is introduced in Section~\ref{sec:proposedm}. In Section~\ref{sec:nr}, a series of Monte-Carlo experiments and an empirical data analysis are performed to evaluate the estimation and classification accuracy of the proposed method. The proposed method is used to analyze empirical data and the results presented in Section~\ref{sec:emp}. Section~\ref{sec:conc} concludes the paper, along with some ideas on how the methodology can be further extended.

\section{Functional logistic regression}\label{sec:FlogitR}

Let us consider an independent and identically distributed random sample $\{Y_i, \X_i(t): i = 1, \ldots, n\}$ from a population $\{Y, \X \}$. The response variable $Y \in \{0,1\}$ takes binary values. The predictor variable $\X = \{\X(t)\}_{t \in \mathcal{T}}$ represents a stochastic process defined on a bounded and closed interval $t \in \mathcal{T}$. This process is assumed to be a random function in the separable Hilbert space $\mathcal{L}_2$, $\mathcal{H}$, the space of square-integrable functions. In FlogitR, the random variable $Y$ conditional on $\X$ is assumed to follow a Bernoulli distribution \citep{ratcliffe2002functional}. Let $\text{Be}(1, \pi_i)$ denote the Bernoulli distribution with success probability $\pi_i = \mathbb{P}[Y_i = 1 \vert \X(t) = \X_i(t)]$, $i = 1, \ldots, n$. Then, the FlogitR presupposes that $Y_i \vert \X_i(t) \sim \text{Be}(1, \pi_i)$, where
\begin{equation}\label{eq:pi}
\pi_i = \frac{\exp \left\{ \beta_0 + \int_{\mathcal{T}} \X(t) \beta(t) dt \right\}}{1 + \exp \left\{ \beta_0 + \int_{\mathcal{T}} \X(t) \beta(t) dt \right\}},
\end{equation}
where $\beta_0 \in \mathbb{R}$ denotes an intercept term and $\beta(t) \in \mathcal{L}_2[\mathcal{T}]$ denotes the regression coefficient function.

To solve~\eqref{eq:pi}, a logit transformation is used, and the FLogitR is expressed as follows:
\begin{equation}\label{eq:logit}
l_i = \ln \left( \frac{\pi_i}{1-\pi_i} \right ) = \beta_0 +\int_{\mathcal{T}} \X_i(t) \beta(t) dt , \quad i = 1, \ldots, n.
\end{equation}
The logit transformation enables understanding of the effects of variables in the model and the way predictions are produced. More specifically, the integral of $\beta(t)$ multiplied by an arbitrary constant, such as $\tau$, can be interpreted as the multiplicative change in the odds of response $Y = 1$ obtained when the functional observation is incremented constantly in  $\tau$ units along $\mathcal{T}$ \citep[see, e.g.,][]{ratcliffe2002functional, escabias2007functional}.

The functional predictor variable in Model~\eqref{eq:logit} is defined in an infinite-dimensional space, while its realizations are practically observed on a discrete set of time points. The direct solution of Model~\eqref{eq:logit} is not appropriately posed. Several approaches, such as pre-determined basis expansion, FPLS, and FPCA have been proposed to overcome this problem. The general idea behind these methods is first to project the infinite-dimensional functional predictor into a finite-dimensional space of basis functions. The infinite-dimensional Model~\eqref{eq:logit} is thus converted to a logit model of $Y$ on the coefficients of the FPLS/FPCA components. Then, the maximum likelihood estimator is used to estimate the regression parameters of the approximate model. This maximum likelihood estimate is multiplied by the basis functions to obtain the estimate of the regression coefficient function $\beta(t)$ \citep[see, e.g.,][for more details about the estimation algorithms of the pre-determined basis expansion, FPLS, and FPCA approaches, respectively]{Ramsay2005, escabias2004principal, escabias2007functional}.

The FPLS and FPCA basis functions are data-driven, and, thus, they are more effective in capturing data-specific patterns, unlike the pre-determined basis functions, such as Fourier, wavelet and B-splines. These data-driven basis functions are frequently favored over pre-determined basis expansion methods in the analysis of functional data. While the dimension-reduction methods lead to effectively estimating the regression coefficient function in the FlogitR, they are susceptible to the presence of outliers. The effects of outliers on the basis functions and their coefficients are reflected in the estimated regression coefficient function obtained from their components. Similarly, the maximum likelihood estimator is also affected by outliers, producing biased estimates for the logit model of $Y$ on the FPLS/FPCA components. Therefore, when outliers are present in the data, the FPLS and FPCA methods may lead to biased parameter estimates and decreased classification accuracy. In this study, by minimizing the effects of outliers, we propose a novel approach to effectively and robustly estimating the regression coefficient of FlogitR in the presence of outliers.

\section{A novel robust method for FlogitR}\label{sec:proposedm}

The proposed method to robustly estimate the FlogitR consists of two steps. First, the RFPCA approach of \cite{bali2011rfpc} is used to compute the principal components and the corresponding principal components scores robustly. With RFPCA, the effects of outliers in the functional predictor are minimized. Second, the parameters of the logistic regression model constructed with a binary response and robust principal component scores are estimated using the weighted \citeauthor{bianco1996robust}'s \citeyearpar{bianco1996robust} (WBY) estimator \citep[see also][]{croux2003implementing}. With the WBY estimator, the effects of outlying observations in both the binary response and RFPCA scores are minimized. Therefore, the proposed method is robust to outliers in both the response and predictor variables.

The RFPCA method proposed by \cite{bali2011rfpc} works similarly to the classical FPCA method but uses a robust M-type scale functional instead of variance, that is the RFPCA decomposition of the functional predictor is obtained as follows:
\begin{equation*}
\X(t) = \sum_{k=1}^K \xi_k \psi_k(t),
\end{equation*}
where $\{\psi_k(t): k = 1, \ldots, K \}$ $\in \mathcal{L}_2(\mathcal{T})$ denotes the orthogonal RFPCA bases and $\{\xi_k: k = 1, \ldots, K \}$ are the corresponding principal component scores. Let $G$ denote the distribution of a given random variable $Z$. Then, the M-scale functional, denoted by $\sigma_M(G)$, is obtained as a solution of
\begin{equation*}
\text{E} \left[ \rho_1 \left( \frac{Z - \mu}{\sigma_M(G)} \right) \right] = \delta,
\end{equation*}
where $\rho_1(\cdot)$ is a loss function, $\mu$ denotes the location parameter, and $\delta = 1/2$. For a given sample $\bm{z} = \left\lbrace z_1, \ldots, z_n \right\rbrace$ with its location estimate $\widehat{\mu}_n$, the M-estimate of scale for the sample $\bm{z}$, denoted by $\widehat{\sigma}_n$, is obtained as solution to the M-estimating equation
\begin{equation}\label{eq:mscale}
\frac{1}{n} \sum_{i=1}^n \rho_1 \left( \frac{z_i - \widehat{\mu}}{\widehat{\sigma}_n} \right) = \delta.
\end{equation}
When $\bm{z}$ is an independently and identically distributed sample from $G$, the M-estimate of scale $\widehat{\sigma}_n$ converges to its true value $\sigma$ \citep[see, e.g.,][]{bali2011rfpc}. In this study and similar to \cite{bali2011rfpc}, the loss-function introduced by \cite{Beaton1974} is considered to compute the M-estimate of scale:
\begin{equation}\label{eq:loss1}
        \rho_{1,c}(u) =
        \left\{ \begin{array}{ll}
            \frac{u^2}{2} \left( 1 - \frac{u^2}{c^2} + \frac{u^4}{3c^4} \right)  & \text{if}~ \vert u \vert \leq c, \\
            \frac{c^2}{6} & \text{if}~ \vert u \vert > c,
        \end{array} \right.
\end{equation}
where the tuning parameter $c$ is used to control the robustness and efficiency of the estimator. In our numerical analyses, we consider $c = 1.56$ and $\delta = 1/2$, leading to an M-scale functional having a 50\% breakdown point at the normal distribution \citep{bali2011rfpc}. Other popular choices of the scale functional are the usual standard deviation and the median absolute deviation. However, the standard deviation has a breakdown point of 0\%; thus, it is not robust to outliers. While the median absolute deviation has a 50\% breakdown point at the normal distribution, its influence function is discontinuous, reflecting some local instability \citep[see, e.g.,][]{bali2011rfpc}.

Let us denote by $\Vert \alpha \Vert^2 = \langle \alpha, \alpha \rangle$ the norm generated by the inner product $\langle \cdot, \cdot \rangle$. Also, let $\mathcal{F}$ and $\mathcal{F}[\alpha]$ denote the distributions of $\X$ and $\langle \alpha, \X \rangle$, respectively.  Then, the orthogonal RFPCA basis for a given M-scale functional $\sigma_M(\mathcal{F})$ is defined as follows
\[ \begin{cases}
\psi_k(\mathcal{F}) = \underset{\begin{subarray}{c}
      \Vert \alpha \Vert = 1
\end{subarray}}{\argmax}~ \sigma_M(\mathcal{F}[\alpha]), & k = 1, \\
\psi_k(\mathcal{F}) = \underset{\begin{subarray}{c}
      \Vert \alpha \Vert = 1, \alpha \in \mathcal{B}_k
\end{subarray}}{\argmax}~ \sigma_M(\mathcal{F}[\alpha]), & k \geq 2,
   \end{cases}
\]
where $\mathcal{B}_k = \left\lbrace \alpha \in \mathcal{L}_2(\mathcal{T}): \langle \alpha, \psi_k ( \mathcal{F} ) \rangle = 0, ~ 1 \leq k \leq \ (k - 1) \right\rbrace$, and the $k$\textsuperscript{th} largest eigenvalue $\lambda_k$ is defined as follows:
\begin{equation*}
\lambda_k(\mathcal{F}) = \sigma_M^2(\mathcal{F}[\psi_k]) = \underset{\begin{subarray}{c}
      \Vert \alpha \Vert = 1, \ \alpha \in \mathcal{B}_k
\end{subarray}}{\max} \sigma_M^2(\mathcal{F}[\alpha]).
\end{equation*}

Let $\sigma_M$ be defined as $\sigma_M(\mathcal{F}_n[\alpha])$. Let $s^2_n: \mathcal{L}_2[\mathcal{T}] \rightarrow \mathbb{R}$ denote the function of the empirical M-scale functional, where $s^2(\alpha) = \sigma_M^2(\mathcal{F}[\alpha])$. In this case, the RFPCA estimates of the orthogonal basis are obtained as follows:
\[ \begin{cases}
\widehat{\psi}_k(t) = \underset{\begin{subarray}{c}
      \Vert \alpha \Vert = 1
\end{subarray}}{\argmax}~ s_n(\alpha), & k = 1, \\
\widehat{\psi}_k(t) = \underset{\begin{subarray}{c}
      \alpha \in \widehat{\mathcal{B}}_k
\end{subarray}}{\argmax}~ s_n(\alpha), & k \geq 2,
   \end{cases}
\]
where $\widehat{\mathcal{B}}_k = \left\lbrace \alpha \in \mathcal{L}_2 (\mathcal{T}): \Vert \alpha \Vert = 1, \langle \alpha, \widehat{\psi}_k \rangle = 0, ~ \forall~ 1 \leq k \leq \ (k - 1) \right\rbrace$. The eigenvalues are computed as follows:
\begin{equation*}
\widehat{\lambda}_k = s^2_n (\widehat{\psi}_k), \quad k \geq 1.
\end{equation*}
Accordingly, the robust principal components are obtained as follows:
\begin{equation*}
\xi_k = \int_{\mathcal{T}} \X(t) \widehat{\psi}_k(t) dt.
\end{equation*}

Based on the RFPCA decomposition of $\X(t)$, the FlogitR model can be expressed in terms of robust principal components as follows:
\begin{equation}\label{eq:apm3}
l_i = \beta_0 + \int_{\mathcal{T}} \X_i(t) \beta(t) dt = \beta_0 + \sum_{k=1}^K \xi_k \gamma_k = \beta_0 + \bm{\Xi}_i^\top \bm{\gamma}, \quad i = 1, 2, \ldots, n,
\end{equation}
where $\gamma_k = \int_{\mathcal{T}} \beta(t) \widehat{\psi}_k(t) dt$, $\bm{\Xi}_i = [\xi_1, \ldots, \xi_K]^\top$, and $\bm{\gamma} = [\gamma_1, \ldots, \gamma_K]^\top$. Let $\bm{Z}_i = (1, \bm{\Xi}_i^\top)^\top$ and $\bm{\theta} = (\beta_0, \bm{\gamma}^\top)^\top$. As shown by \cite{croux2003implementing}, a robust estimate of $\bm{\theta}$ can be obtained as follows:
\begin{equation}\label{eq:objf}
\widehat{\bm{\theta}} = \argmin_{\bm{\theta}} \sum_{i=1}^n \rho_2(\bm{Z}_i^\top \bm{\theta}; Y_i),
\end{equation}
where $\rho_2(\cdot)$ is a positive and differentiable function everywhere. For any given score value $\kappa = \bm{Z}_i^\top \bm{\theta}$, it is required that $\rho_2(\kappa; 0) = \rho_2(-\kappa; 1)$. Without loss of generality, let us consider a univariate and decreasing function $\rho_2(\kappa) = \rho_2(\kappa; 0)$ satisfying $\lim_{\kappa \rightarrow - \infty}~ \rho_2(\kappa) = 0$. Here, the value of $\rho_2(\kappa)$ for an observation corresponding to $Y = 0$ shows the effect of the score $\kappa$ on the value of the objective function given in~\eqref{eq:objf}.

From~\eqref{eq:objf}, $\widehat{\bm{\theta}}$ is an M-type estimator, and its first-order condition is given as follows:
\begin{equation*}
\frac{1}{n} \sum_{i=1}^n \rho_2^{\prime}(\bm{Z}_i^\top \bm{\theta}; Y_i) \bm{Z}_i^\top = 0,
\end{equation*}
where $\rho_2^{\prime}(\kappa;0) = \partial \rho_2(\kappa; 0) / \partial \kappa$, which satisfies $\rho_2^{\prime}(\kappa;1) = - \rho_2^{\prime}(-\kappa;0)$, represents the derivative of the loss function $\rho_2$. The estimation strategy proposed by \cite{croux2003implementing} takes into account the WBY estimator:
\begin{equation}\label{eq:byest}
\widehat{\bm{\theta}} = (\widehat{\beta}_0, \widehat{\bm{\gamma}}^\top)^\top =  \argmin_{\bm{\theta}} \sum_{i=1}^n \omega_i~ \{ \rho_2[d(\bm{Z}_i^\top \bm{\theta}; Y_i)] + C(\bm{Z}_i^\top \bm{\theta}) \},
\end{equation}
where $d(\bm{Z}_i^\top \bm{\theta}; Y_i)$ is the deviance component, which is obtained as follows:
\begin{equation*}
d(\bm{Z}_i^\top \bm{\theta}; Y_i) = - Y_i \ln F(\bm{Z}_i^\top \bm{\theta}) - (1 - Y_i) \ln[1 - F(\bm{Z}_i^\top \bm{\theta})],
\end{equation*}
where $F(u) = 1 / [1 + \exp(-u)]$ denotes the logit model function, $C(\bm{Z}_i^\top \bm{\theta})$ is the bias correction term, which is computed as follows:
\begin{equation*}
C(\kappa) = D[F(\kappa)] + D[1 - F(\kappa)] + D(1),
\end{equation*}
where $D(u) = \int_0^u \rho_2^{\prime}(- \ln u) du$, and $\omega_i$ denotes the weight computed using robust Mahalanobis distance.

From~\eqref{eq:byest}, the WBY estimator depends on the choice of the loss function $\rho_2$. In this study, similar to \cite{croux2003implementing}, the following loss function is considered:
\begin{align*}
\rho_2(u)=
\begin{cases}
u e^{-\sqrt{c}}, & \text{if}\ u \leq c, \\
-2e^{-\sqrt{u}}(1+\sqrt{u})+e^{-\sqrt{c}}(2(1+\sqrt{c})+c), & \text{otherwise},
\end{cases}
\end{align*}
where $c$ is the tuning parameter that controls the robustness and efficiency of the estimator. As the value of $c$ increases, the estimator's efficiency increases, and vice versa. In this study, $c$ is chosen to be~$0.5$ as suggested by \cite{croux2003implementing}. 

The estimate of the regression coefficient function of the FlogitR in~\eqref{eq:logit} is obtained as follows:
\begin{equation}\label{eq:propest}
\widehat{\beta}(t) = \sum_{k=1}^K \widehat{\psi}_k(t)  \widehat{\gamma}_k.
\end{equation}
If the functional predictor is contaminated by outliers, the effect of these observations is minimized, since the principal components and their associated scores are obtained robustly by the RFPCA method. By the WBY estimator, the parameters of the logistic regression model constructed using binary response and robust principal component scores, $ \widehat{\bm{\gamma}}$, are robustly estimated. Therefore, the estimate of $\beta(t)$ obtained by the proposed method in~\eqref{eq:propest} is robust to outliers in both the response and predictor variables. The asymptotic consistency and influence function of $\widehat{\beta}(t)$ are discussed in Appendix.

\subsection{Computation details}

The orthogonal bases in the FPCA or RFPCA methods, $\{\psi_k(t): k = 1, \ldots, K \}$, are unknown in practice, and they are approximated via a suitable general basis expansion method (the same procedure is also applied in the FPLS-based methods). In this study, the B-spline basis expansion method is employed to approximate the orthogonal bases of FPCA and RFPCA approaches. Additionally, for both methods, the functional predictor is centered. 

Let us assume that the elements of the functional predictor, $\{\X_i(t):~i = 1, \ldots, n \}$, are generated by a set of B-spline basis functions $\{\phi_m(t):~m = 1, \ldots, M\}$ as follows:
\begin{equation*}
\X_i(t) = \sum_{m=1}^M a_{iM} \phi_m(t), \quad i = 1, \ldots, n,
\end{equation*}
where $\{a_{im}:~m = 1, \ldots, M\}$ are the B-splines basis expansion coefficients. Let $\mu_{\X}(t)$ denote the mean curve for the functional predictor. Note that for the FPCA approach, the classical mean, i.e., $\mu_{\X}(t) = \frac{1}{n}\sum_{i=1}^n \X_i(t)$, is used to center the functional predictor, while the $\text{L}_1-\text{Median}$, i.e., $\mu_{\X}(t) = \argmin_{u \in \mathcal{H}} \sum_{i=1}^n \Vert \X_i(t) - u \Vert$, is used in the RFPCA approach \citep[see, e.g.,][]{bali2011rfpc}. Then, the functional principal components are computed as follows:
\begin{equation*}
\xi_k = \int_{\mathcal{T}} [\X(t) - \mu_{\X}(t)] \widehat{\psi}_k(t) dt,
\end{equation*}
where
\begin{equation*}
\widehat{\psi}_k(t) = \sum_{m = 1}^M \psi_{km} \phi_m(t), \quad k = 1, \ldots, K,~~ m = 1, \ldots, M,
\end{equation*}
and $\{\psi_{km}:~m = 1, \ldots, M\}$ denotes the B-splines basis expansion coefficients.

From the expression given above, the B-splines basis expansion coefficients $\{\psi_k:~k = 1, \ldots, K\}$ are the eigenvectors of $\bm{A} \bm{\Phi}^{1/2}$, where $\bm{A}$ is the matrix of basis expansion coefficients $\{a_{im}:~m = 1, \ldots, M\}$ in its columns, and $\bm{\Phi}^{1/2}$ is the square root of the inner-product matrix $\bm{\Phi} = \int_{\mathcal{T}} \Phi(t) \Phi^\top(t) , dt$ with $\Phi(t) = [\phi_1(t), \ldots, \phi_M(t)]^\top$ \citep[see, e.g.,][for more information]{escabias2004principal, Ramsay2005, ocana2007}.

Following the above definitions, the performance of the proposed method depends on two factors: the number of B-splines basis expansion functions $M$ and the number of principal components $K$. Several criteria, such as the generalized Bayesian information criterion \citep{konishi2004}, generalized information criterion \citep{konishi2008}, modified Akaike information criterion \citep{fujikoshi1997}, and generalized cross-validation \citep{craven1979}, have been proposed to determine the optimum number of basis expansion functions $M$ \citep[see also][for a comparison of these criteria]{BS2022}. However, these methods may be time-consuming. For simplicity, in our numerical analyses, the optimum number for $M$ is determined by minimizing the square of the residuals of the observed and predicted values of the functional predictor as follows:
\begin{equation*}
\varphi^2 = \frac{1}{n - M} \sum_{i = 1}^n \int_{\mathcal{T}} [\X_i(t) - \widehat{\X}_i(t)]^2 , dt,
\end{equation*}
where $\widehat{\X}_i(t)$ is the predicted curve using $M$ -- B-splines basis expansion functions. 

For this purpose, we perform a standard grid search with a set of candidate values for $M$, $\{M = 4, \ldots, \min(40, J/4)\}$, where $J$ is the number of observations on the predictor curves (we use quadratic B-splines basis expansion functions, the $M$ should start from 4). Large values of $M$ will result in less sum of squared error but at the expense of complicating the model. In the grid search, we determine the optimum value of $M$ if two consecutive $\varphi^2$ values are less than $10^{-6}$. This approach is more appropriate for curves without measurement errors. However, if the curves $\X_i(t)$ are observed with measurement errors, a pre-smoothing step of the curves may be necessary before computing $\varphi^2$. For the optimum number of principal components $K$, we consider the explaining variance approach, where the number of functional principal components is determined as optimum by explaining at least 99\% explaining variation in the data.

\section{Monte-Carlo simulation}\label{sec:nr}

The finite-sample performance of the proposed RFPCA method in terms of estimation, classification, and computing time is evaluated through a series of Monte-Carlo experiments and an empirical data example, hand radiography data. The results obtained by the proposed method are compared with those of classical FPLS and FPCA, as well as the RFPLS method. 

Throughout the experiments, 200 Monte-Carlo runs are performed, and the following process is repeated. First, a dataset consisting of $n=1000$ observations is generated. The generated dataset is divided into two parts, a training sample consisting of 700 randomly selected observations and a test sample consisting of the remaining 300 observations. The FPCA, FPLS, RFPLS, and RFPCA methods are fitted on the training set. Depending on the fitted models, the values of the binary response variable in the test sample are predicted. 

\subsection{Monte-Carlo simulations: Estimation and classification accuracy}

Two different scenarios are considered for the data generation process. First, the dataset is generated from a smooth data generation process, and there are no clear outliers in the data. This scenario aims to test the consistency of the proposed method in terms of parameter estimation and classification of the response variable. Second, $700 \times [1\%, 5\%, 10\%, 20\%]$ of the observations randomly selected from the training sample are contaminated by outliers. This scenario aims to evaluate the robustness of the proposed method. In addition, the computational efficiency of the proposed method is compared with existing methods.

For the FPLS and RFPLS methods, we consider the iterative algorithms proposed by \cite{escabias2007functional} and \cite{mutis2022robust}. In these algorithms, the latent components are extracted based on the usual Wald test, and the algorithms stop at the step where the latent component is not significant in the Wald test. The estimation performance of the methods is compared using the integrated mean squared error (IMSE):
\begin{equation*}
\text{IMSE} = \int _{\mathcal{T}}\left [ \beta(t)-\widehat{\beta}(t) \right ]^{2}dt.
\end{equation*}
The classification performance of the methods is compared using the area under the receiver operating characteristic curve (AUC). To compare the estimation and predictive performance of the methods, we calculate the median and median absolute deviation of the IMSE and AUC values computed over 200 Monte-Carlo runs. The numerical analyses are performed on a Huawei D16 (2022) PC with a $12$\textsuperscript{th} Generation Intel(R) Core(TM) i5-12450H, 2.00 GHz, and 16 GB RAM in performance mode. An online supplement file provides an example \Rlogo\ script for all the methods considered in this study.

In the data generation process, the realizations of the functional predictor variable are generated in 201 equally spaced points in the interval $\mathcal{T}= [0,1 ]$. The following process is used to generate the values of the functional predictor variable:
\begin{equation*}
\X_i(t) = \sum_{l = 1}^5 \zeta_l \psi_l(t),
\end{equation*}
where $\zeta_l \sim \mathcal{N}(0, 4 l^{-3/2})$ and $\psi_l(t) = e^{-l^2 t} + \sin(l \pi t)$. The regression coefficient function is chosen as $\beta(t) =\sin (\pi t)$, and the values of the binary response variable are then obtained as a linear combination of the predictor variable and the regression coefficient function as follows:
\begin{equation*}
l_i = \int_{0}^{1}\mathcal{X}_{i}(t)\beta(t)dt, \quad i = 1, 2, \ldots, 1000.
\end{equation*}

\begin{figure}[!htb]
\centering
\includegraphics[width=5.93cm]{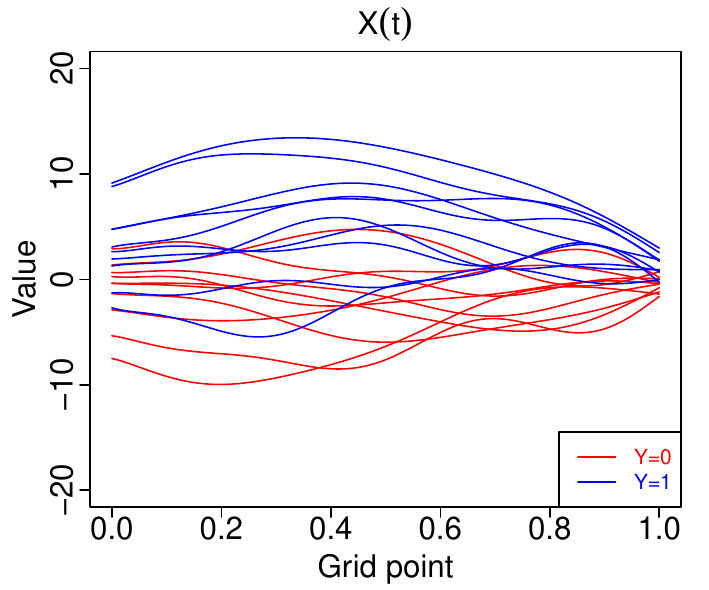}
\includegraphics[width=5.93cm]{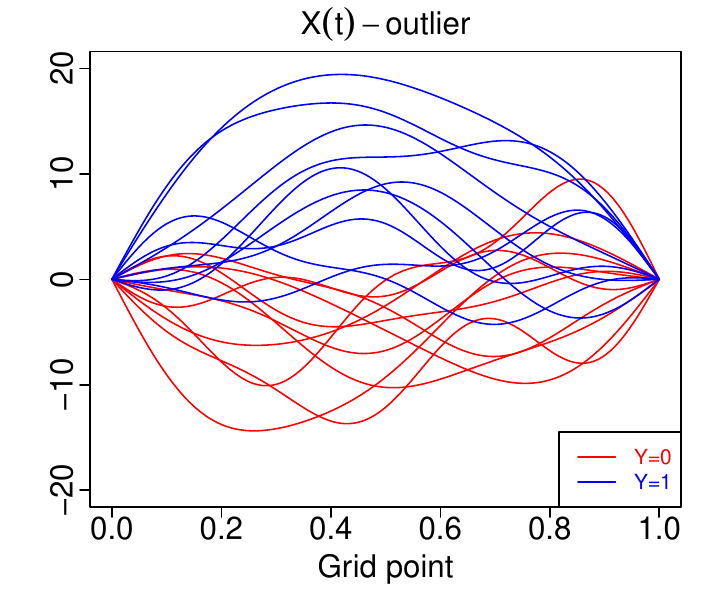}
\includegraphics[width=5.93cm]{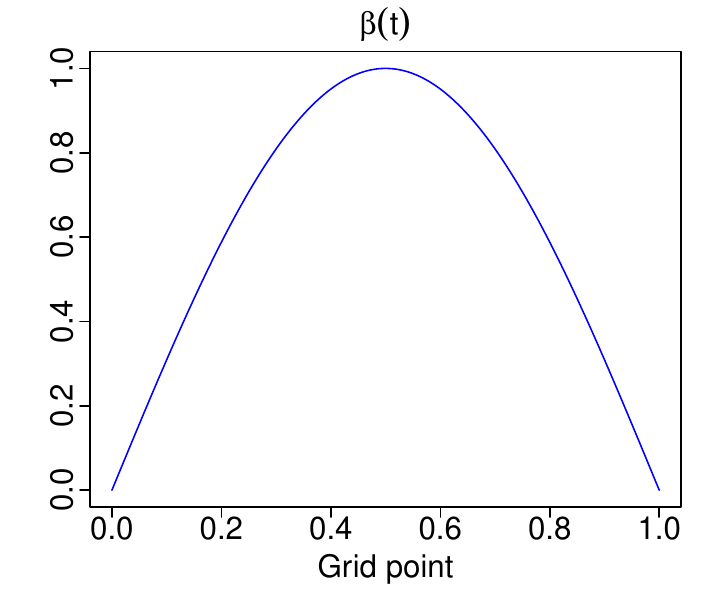}
\caption{\small{Graphical display of the generated data and the regression coefficient function. The elements of the functional predictor generated under the first scenario are presented in the left panel, while those generated under the second scenario are presented in the middle panel. The regression coefficient function is presented in the right panel}.}\label{fig:Fig_1}
\end{figure}

We consider two scenarios where data are outlier-free and contaminated with outliers. In the latter case, $700 \times [1\%, 5\%, 10\%, 20\%]$ of randomly selected observations from the training sample are replaced with outlying observations. The outlying observations in the functional predictor are generated as $\widetilde{\X}_i = 1.25 \sum_{l = 1}^5 \zeta_l \widetilde{\psi}_l(t)$ where $\widetilde{\psi}_l(t) = 2 \sin(l \pi t)$. In addition, the corresponding observations of the binary response variable are replaced by $\widetilde{Y}_i = 1 - Y_i$; that is, the values of the response variable corresponding to the outlying functional predictor's observations are changed from 1 to 0 and vice versa. Therefore, the response and predictor variables are contaminated by outlying observations. The proposed method is expected to produce improved estimation and classification performance compared with the classical methods and produce similar results to the RFPLS method with less computing time. A graphical display of the generated data and the regression coefficient function is presented in Figure~\ref{fig:Fig_1}. 

From the middle panel of Figure~\ref{fig:Fig_1}, some observations are inside the range $[-10, 10]$ as is the outlier-free case (left panel), in which there are shape outliers that undermine the non-robust methods. The observations outside the range $[-10, 10]$ are both shape and magnitude outliers, and they negatively impact the estimation methods to a greater extent than those of shape outliers.

\begin{center}
\tabcolsep 0.23in
\begin{longtable}{@{}llccccc@{}} 
\caption{Computed median IMSE, AUC, and their MAD values (given in brackets) over 200 simulation-runs for different contamination levels (CLs). 0\% CL corresponds to outlier-free data generation case.}\label{tab:tab_1} \\
\toprule
{Metric} & {Method} & CL = 0\% & CL = 1\% & CL = 5\% & CL = 10\% & CL = 20\% \\
\midrule
\endfirsthead
\toprule
{Metric} & {Method} & CL = 0\% & CL = 1\% & CL = 5\% & CL = 10\% & CL = 20\% \\
\midrule
\endhead
\midrule
\multicolumn{7}{r}{Continued on next page} \\ 
\endfoot
\endlastfoot
\multirow{8}{*}{IMSE} & FPLS & 0.241 & 0.290 & 0.817 & 1.938 & 1.958 \\
& & (0.023) & (0.070) & (0.501) & (1.213) & (0.687) \\
& FPCA & \textbf{0.041} & 0.088 & 0.709 & 1.729 & 2.365 \\
& & (0.032) & (0.080) & (0.363) & (0.631) & (0.621) \\
& RFPLS & 0.320 & 0.340 & 0.478 & 0.665 & 0.811 \\
& & (0.031) & (0.023) & (0.145) & (0.279) & (0.242) \\
& RFPCA & 0.049 & \textbf{0.051} & \textbf{0.076} & \textbf{0.131} & \textbf{0.429}  \\
& & (0.043) & (0.044) & (0.059) & (0.100) & (0.248) \\
\cmidrule(l){2-7}
\multirow{8}{*}{AUC} & FPLS & \textbf{0.856} & 0.855 & 0.821 & 0.792 & 0.749 \\
& & (0.021) & (0.018) & (0.025) & (0.031) & (0.039) \\
& FPCA & \textbf{0.856} & 0.857 & 0.841 & 0.816 & 0.783 \\
& & (0.019) & (0.017) & (0.020) & (0.022) & (0.023) \\
& RFPLS & 0.828 & 0.826 & 0.804 & 0.777 & 0.739 \\
& & (0.024) & (0.022) & (0.027) & (0.031) & (0.035) \\
& RFPCA & \textbf{0.856} & \textbf{0.858} & \textbf{0.858} & \textbf{0.856} & \textbf{0.846} \\
& & (0.019) & (0.016) & (0.020) & (0.020) & (0.023) \\
\bottomrule
\end{longtable}
\end{center}

The results are presented in Table~\ref{tab:tab_1}, which indicates that, irrespective of the presence of outliers and contamination levels, the FPCA-based methods yield improved estimation and classification accuracy results compared to their FPLS-based counterparts. When no outliers are present in the data, all methods (except RFPLS) exhibit similar classification accuracy (similar AUC values). In contrast, in this scenario, FPCA and RFPCA yield significantly improved IMSE values compared to FPLS and RFPLS. The RFLS method, in this case, yields the worst IMSE values. When the generated data are contaminated by outliers, all methods, except the proposed RFPCA, yield poor classification results. Conversely, RFPCA minimizes the effects of outliers and produces AUC values similar to those obtained in the outlier-free scenario. In other words, while the classification accuracy of existing methods decreases in the presence of outliers, the proposed RFPCA method yields almost similar and consistent AUC values for all contamination levels.

From Table~\ref{tab:tab_1}, it is evident that all methods are affected by outliers, resulting in larger IMSE values compared to those obtained in the outlier-free scenario. However, compared with non-robust FPLS and FPCA methods, the effect of outliers on the estimation performance of RFPLS and RFPCA is limited. The robust methods yield significantly improved IMSE values than the non-robust methods in the presence of outliers. For example, FPLS produces 8.23 times more IMSE values when the contamination level is 20\% than its IMSE value under the outlier-free data generation case, while this increment is 2.53 for RFPLS. Similarly, FPCA produces 57.68 times more IMSE values when the contamination level is 20\% than its IMSE value under the outlier-free data generation case, while this increment is 8.75 for the proposed RFPCA.

\begin{figure}[!htb]
\centering
\includegraphics[width=4.42cm]{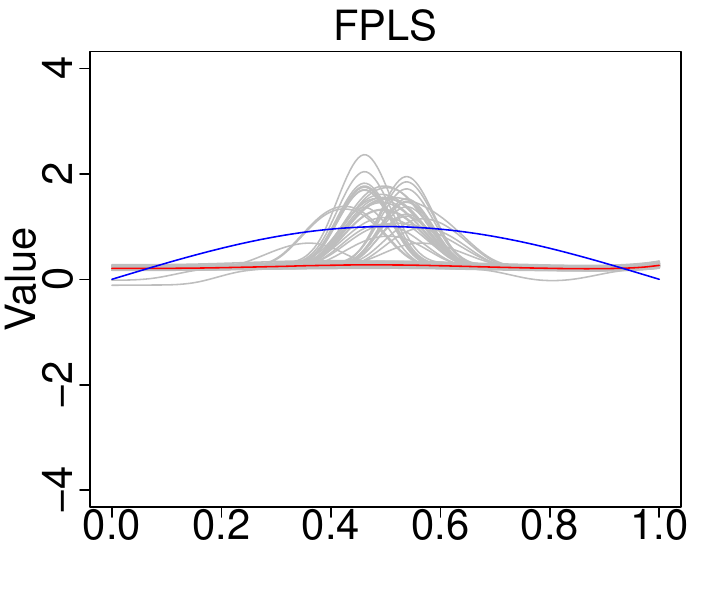}
\includegraphics[width=4.42cm]{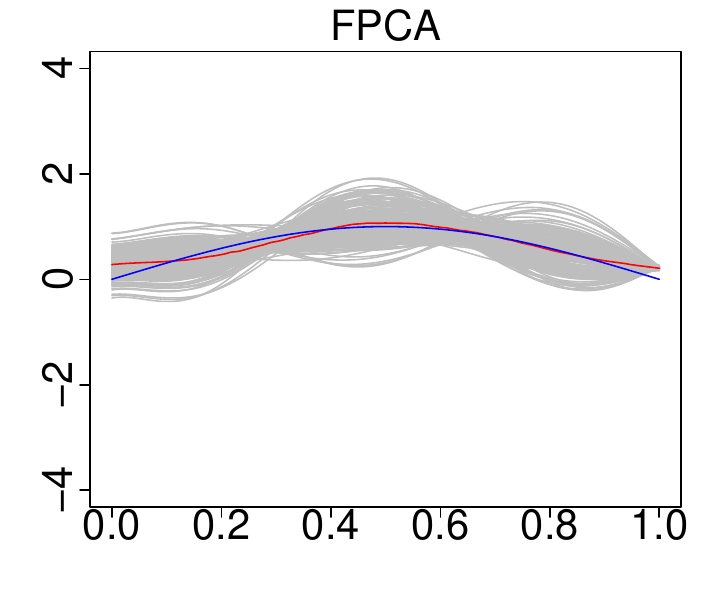}
\includegraphics[width=4.42cm]{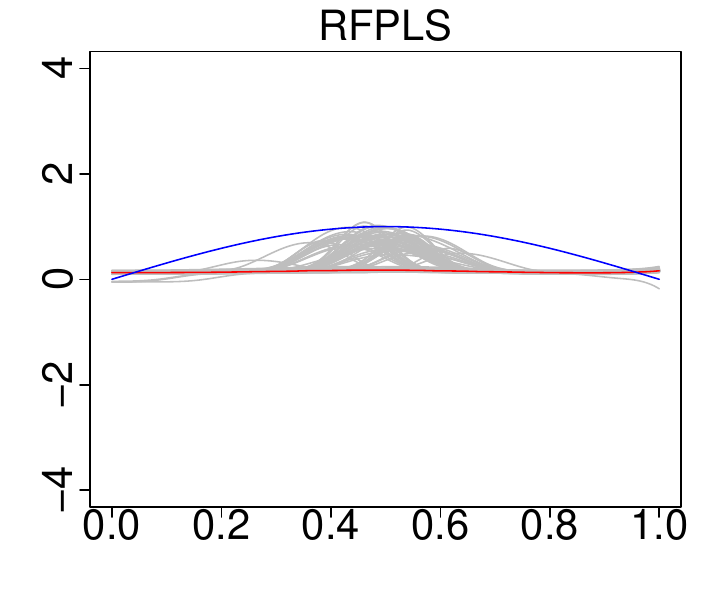}
\includegraphics[width=4.42cm]{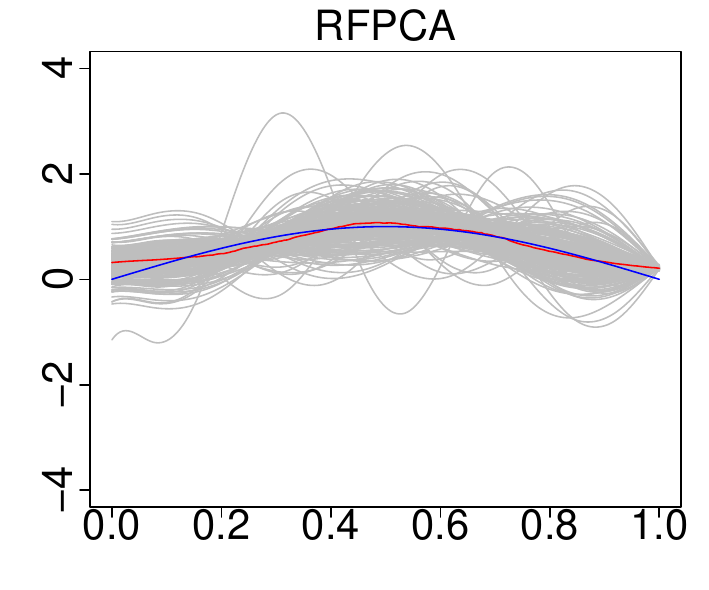}
\\ 
\includegraphics[width=4.42cm]{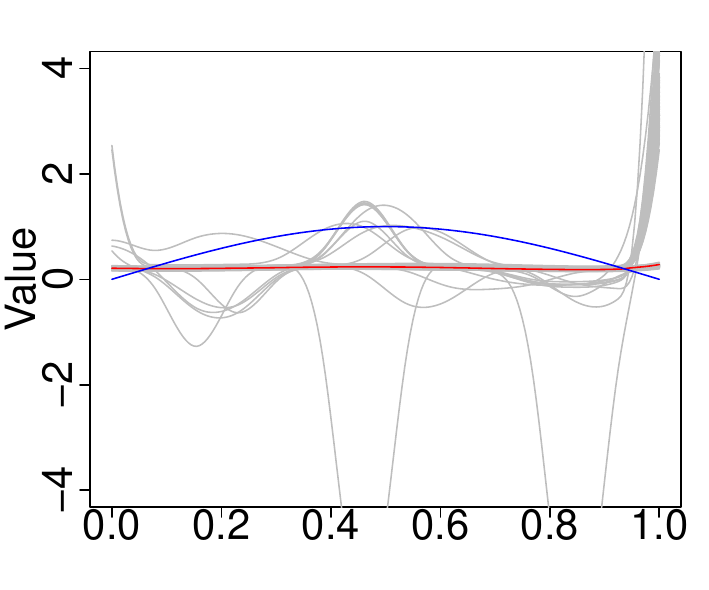}
\includegraphics[width=4.42cm]{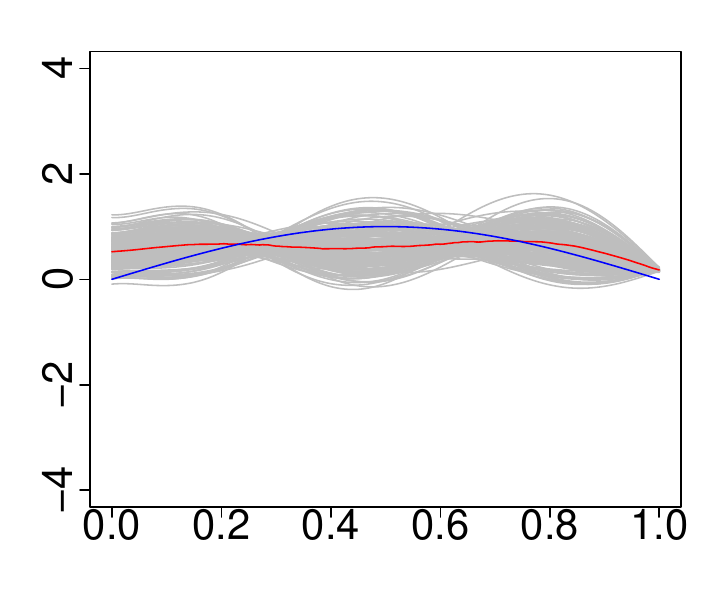}
\includegraphics[width=4.42cm]{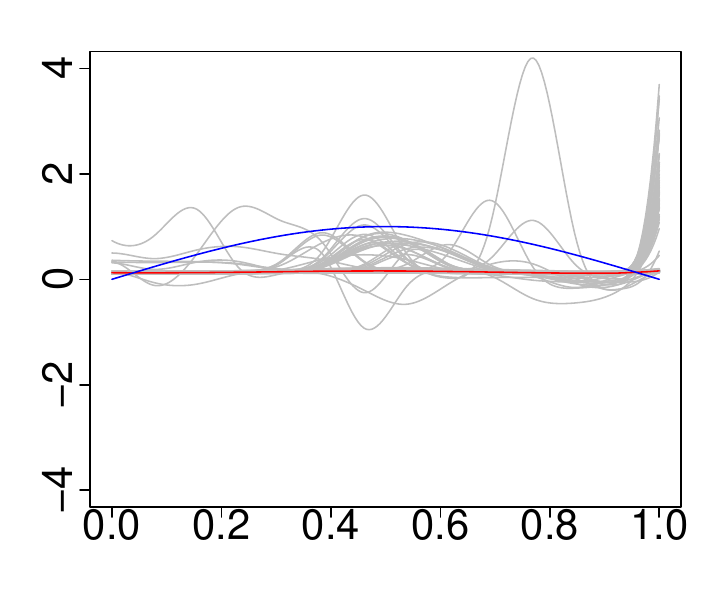}
\includegraphics[width=4.42cm]{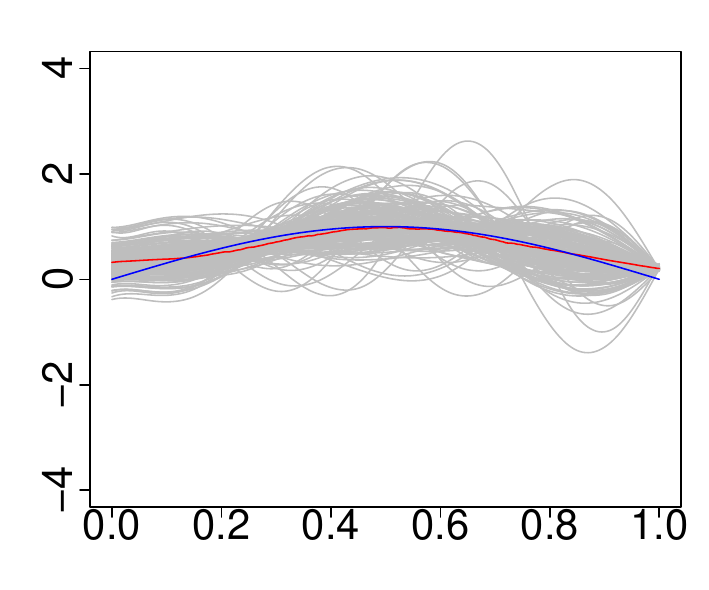}
\\   
\includegraphics[width=4.42cm]{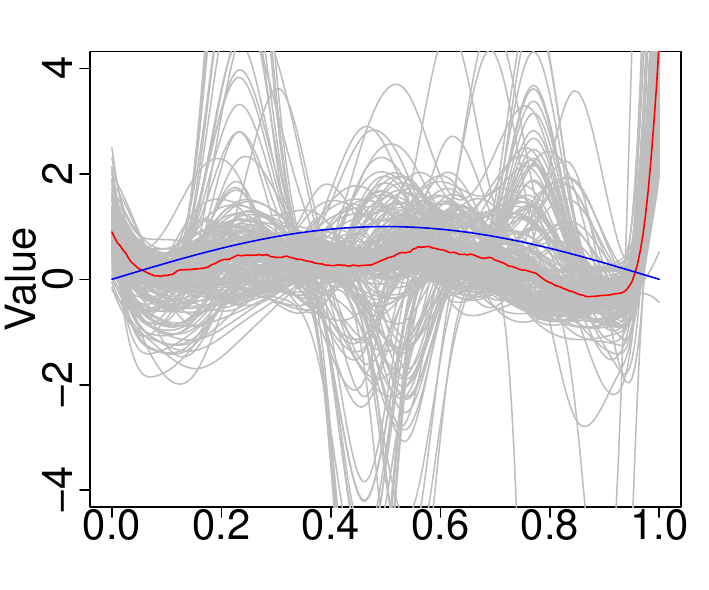}
\includegraphics[width=4.42cm]{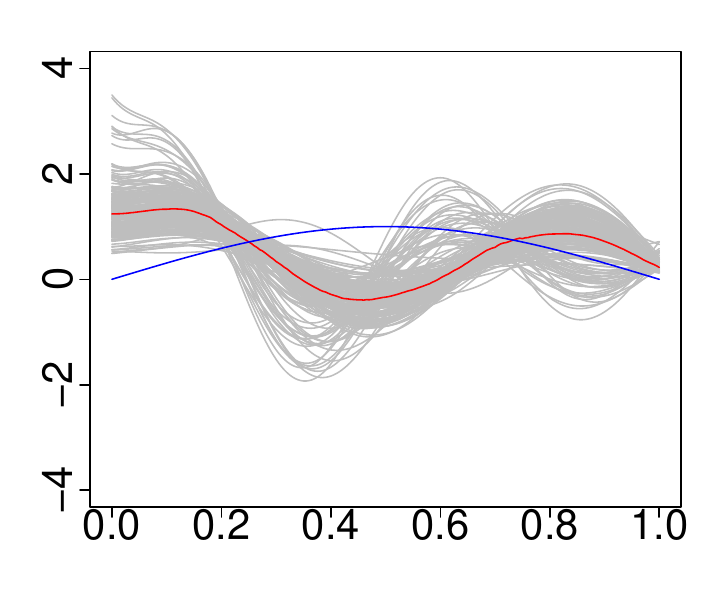}
\includegraphics[width=4.42cm]{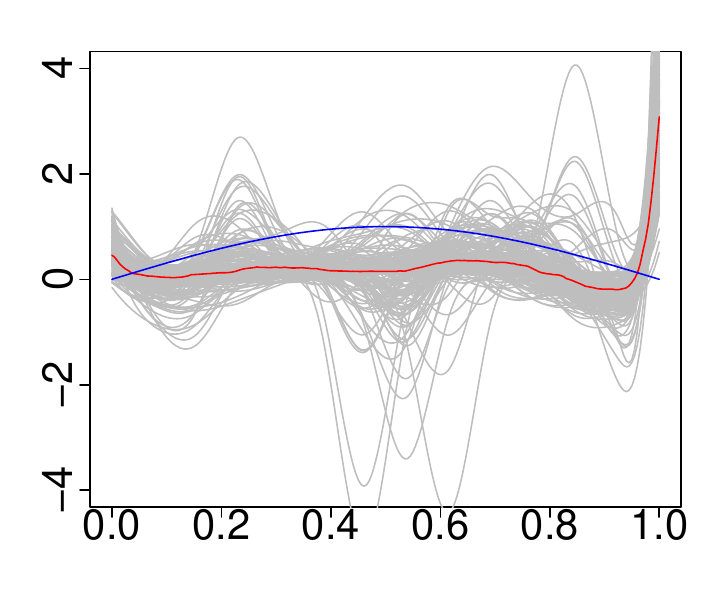}
\includegraphics[width=4.42cm]{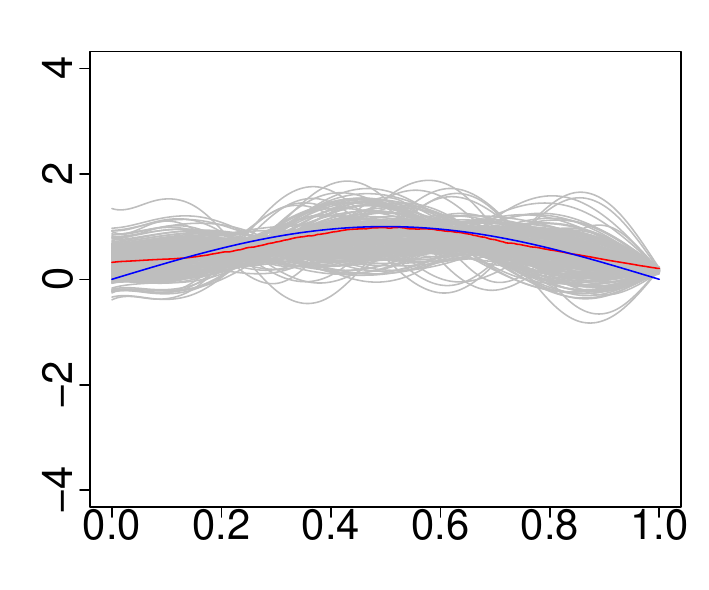}
\\ 
\includegraphics[width=4.42cm]{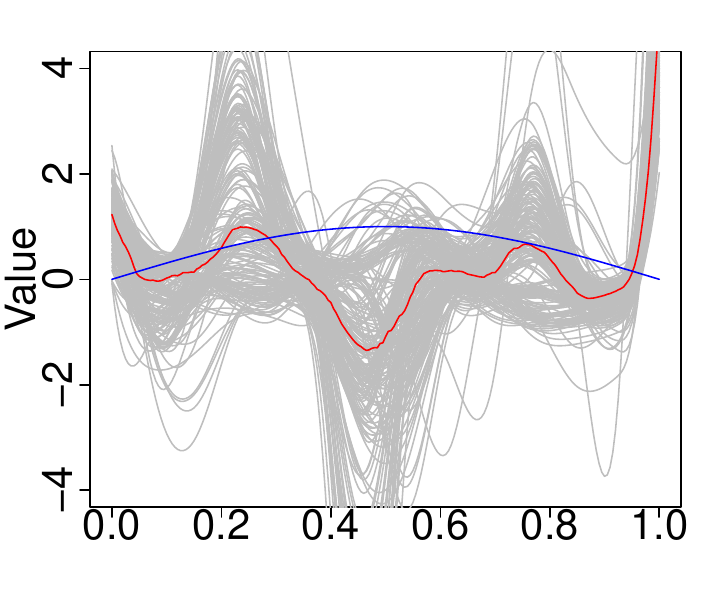}
\includegraphics[width=4.42cm]{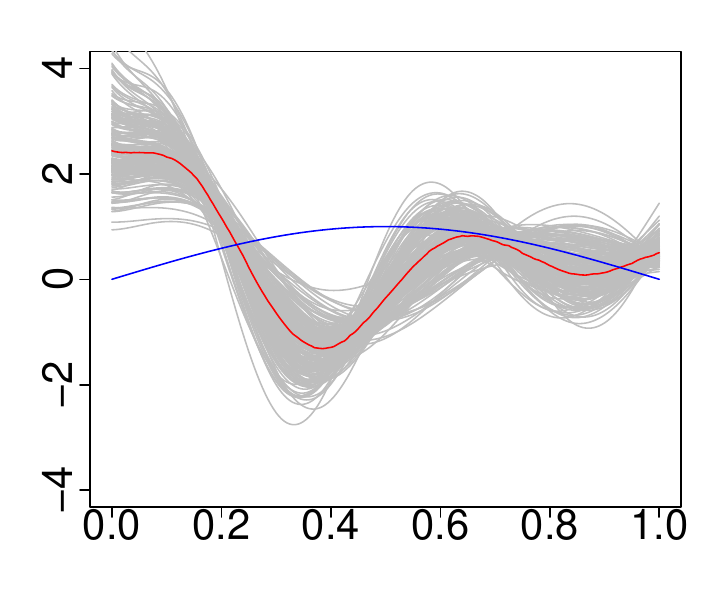}
\includegraphics[width=4.42cm]{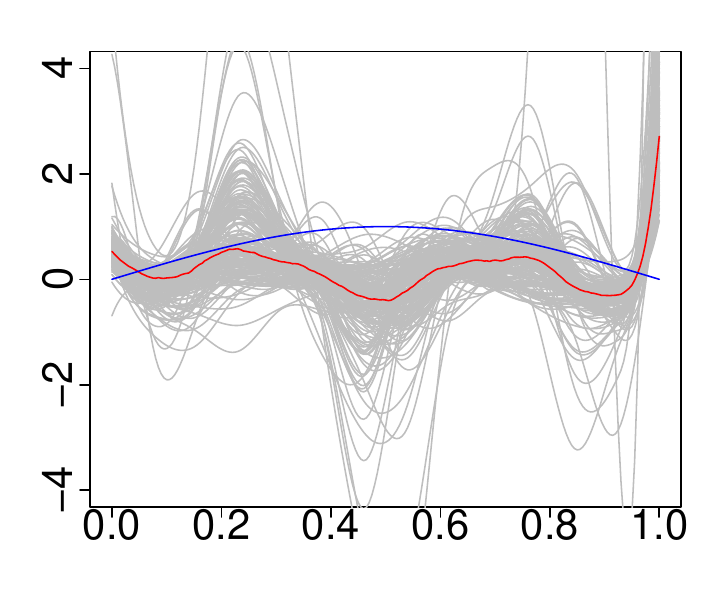}
\includegraphics[width=4.42cm]{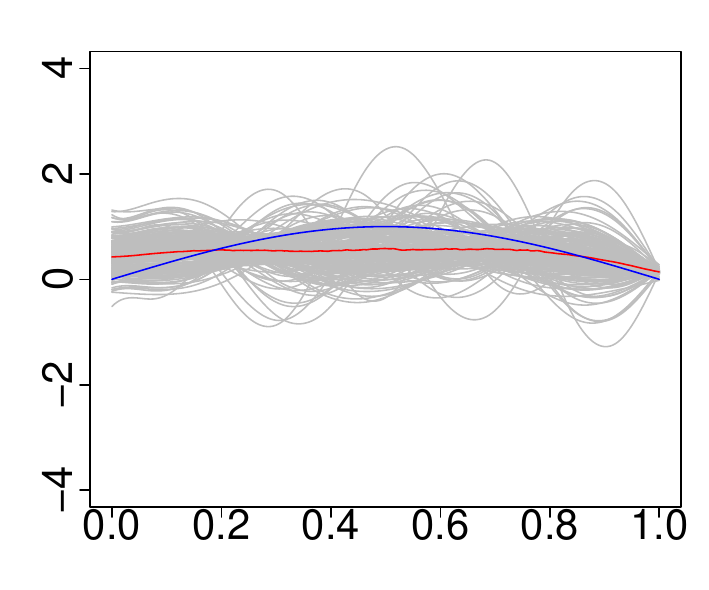}
\\ 
\includegraphics[width=4.42cm]{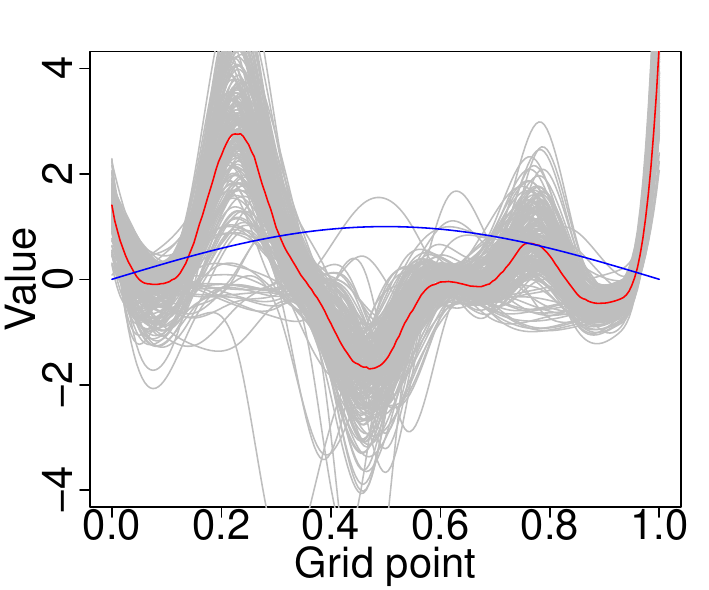}
\includegraphics[width=4.42cm]{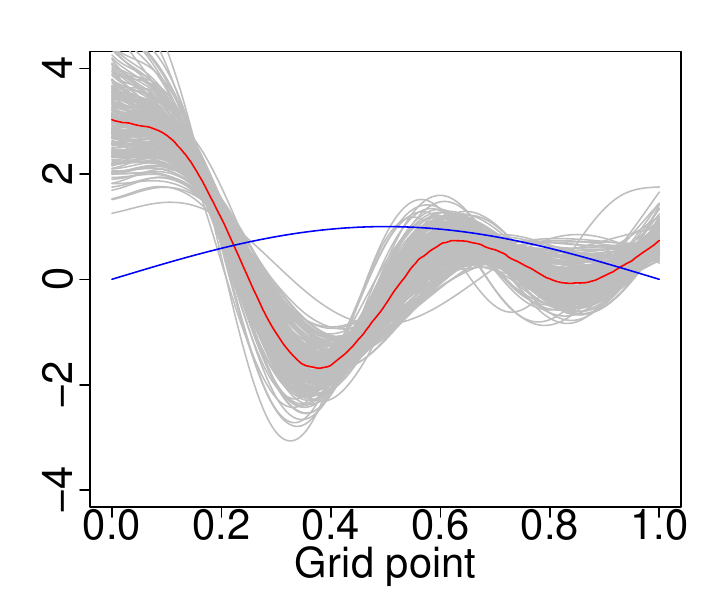}
\includegraphics[width=4.42cm]{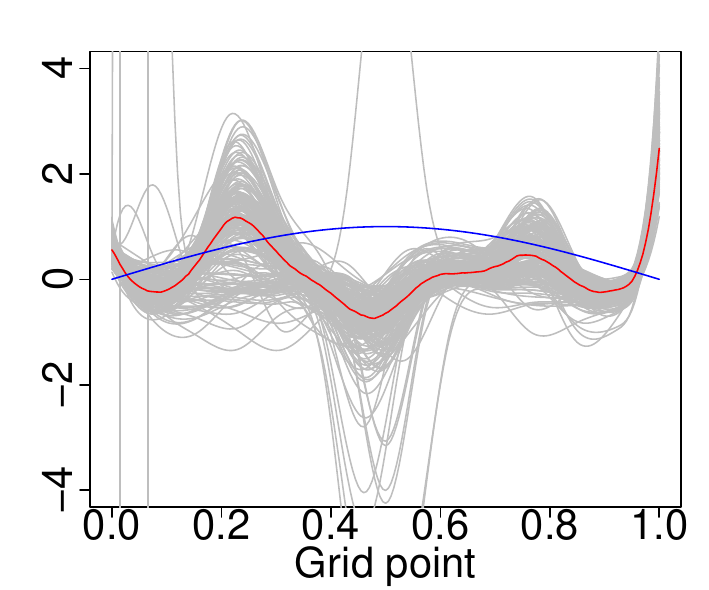}
\includegraphics[width=4.42cm]{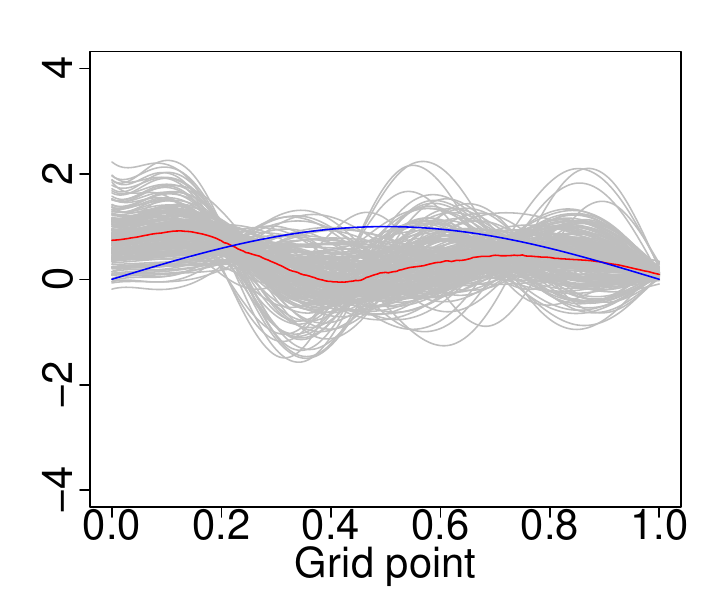}
\caption{\small{Graphical display of the parameter estimates (grey curves), their medians (red curves), and true parameter functions (blue curves) for the FPLS (first column), FPCA (second column), RFPLS (third column), and RFPCA (fourth column) methods. The results obtained when no outliers are present in the data are presented in the first row, and the results obtained when $[1\%, 5\%, 10\%, 20\%]$ of the training sample are presented in the second, third, fourth, and fifth rows, respectively}.}\label{fig:Fig_2}
\end{figure}

A graphical display of the parameter estimates obtained from 200 Monte-Carlo experiments, their medians, and the true parameter functions is presented in Figure~\ref{fig:Fig_2}. The results in this figure also support the findings presented in Table~\ref{tab:tab_1}; that is, the proposed method produces more stable parameter estimates, similar to the true parameter function for all contamination levels.

From our results, the FPLS-based methods generally worse results than the FPCA-based methods, while they are expected to produce similar results. This difference may be because of the iterative algorithm used in the FPLS method, which extracts the latent components and their coefficients using the usual Wald test and stops at the step where the latent component is insignificant in the Wald test. With this algorithm, the number of FPLS components less than it should be may be used in the estimation step (or vice versa), leading to poor estimation and classification results. The performance of the FPLS-based methods may be increased using a large number of components resulting in at least 99\% explained variation in the data (as in the FPCA-based methods).

\subsection{Monte-Carlo simulations: Computing time}

We compare the computational efficiencies of the methods through a series of Monte-Carlo experiments. For this purpose, two different scenarios are considered. In the first scenario, the effects of the number of basis expansion functions on the computing performance of the methods are investigated by generating $n = 1000$ observations and using $M = [5, 10, 15, 20, 25, 30]$ B-spline basis expansion functions. Note that the computing times are obtained from a single Monte Carlo experiment. In the second scenario, the effects of the number of observations on the computing performance of the methods are investigated by using fixed $M = 15$ B-spline basis expansion functions and $n = [100, 300, 500, 1000, 1500, 2000]$ observations. The computing times obtained under both scenarios are presented in Table~\ref{tab:Tab_1}.

From~Table~\ref{tab:Tab_1}, it is observed that non-robust FPLS and FPCA methods require less computing time than those of robust methods. It is also observed that the proposed RFPCA method requires less than 1 second to produce the results for all considered $M$ values, while the RFPLS method requires 29--151 times more computing time than RFPCA. Table~\ref{tab:Tab_1} also indicates that, for different sample sizes, non-robust methods require much less computation time than robust methods, and the proposed RFPCA method requires 20--103 times less computing time than RFPLS.

\begin{table}[!htb]
\centering
\tabcolsep 0.17in
   \caption{\small{Elapsed computing times for the FPLS, FPCA, RFPLS, and RFPCA methods (in seconds). The computing times are obtained from a single Monte-Carlo experiment when $n = 1000$ for $M = [5,10,15,20,25,30]$} (first five columns) and $M = 15$ for $n = [100,300,500,1000,1500,2000]$ (last five columns).}
   \label{tab:Tab_1}
\begin{tabular}{@{}rcccc|rcccc@{}}
\toprule
$M$ & FPLS & FPCA & RFPLS & RFPCA & $n$ & FPLS & FPCA & RFPLS & RFPCA\\
\midrule
5 	& 0.08 & 0.02 & 14.62 & 0.50 & 100 & 0.03 & 0.02 & 3.09  & 0.15 \\
10 	& 0.17 & 0.04 & 18.39 & 0.39 & 300 & 0.12 & 0.03 & 8.29  & 0.34 \\
15 	& 0.14 & 0.03 & 25.16 & 0.32 & 500 & 0.20 & 0.03 & 22.29 & 0.48 \\
20 	& 0.41 & 0.03 & 41.49 & 0.50 & 1000 & 0.15 & 0.03 & 35.12 & 0.34 \\
25 	& 0.56 & 0.05 & 66.51 & 0.44 & 1500 & 0.19 & 0.03 & 52.39 & 1.39 \\
30 	& 0.73 & 0.07 & 92.10 & 0.61 & 2000 & 0.36 & 0.05 & 97.20 & 1.47 \\
\bottomrule
\end{tabular}
\end{table}

\section{Analysis of hand radiograph data}\label{sec:emp}

\begin{figure}[!htb]
\centering
\includegraphics[width=8.9cm]{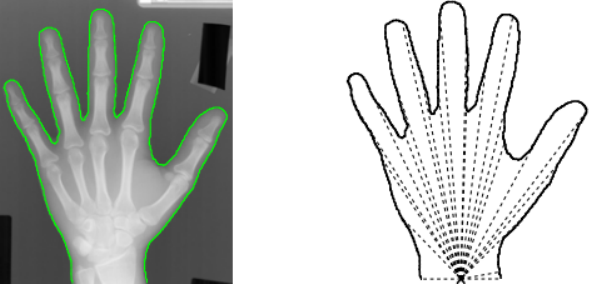}
\includegraphics[width=8.9cm]{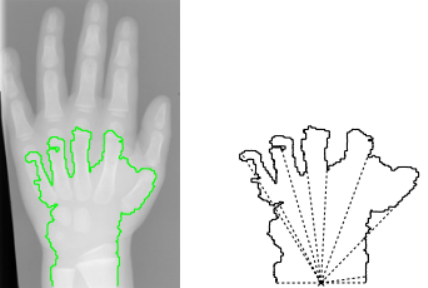}
\caption{\small{Graphical display of a correctly segmented hand outline from a radiograph (first panel), its 1-D series (second panel), an incorrectly located hand outline from a radiograph (third panel), and its 1-D series (fourth panel)}.}\label{fig:Fig_3}
\end{figure}

\begin{figure}[!htb]
\centering
\includegraphics[width=5.93cm]{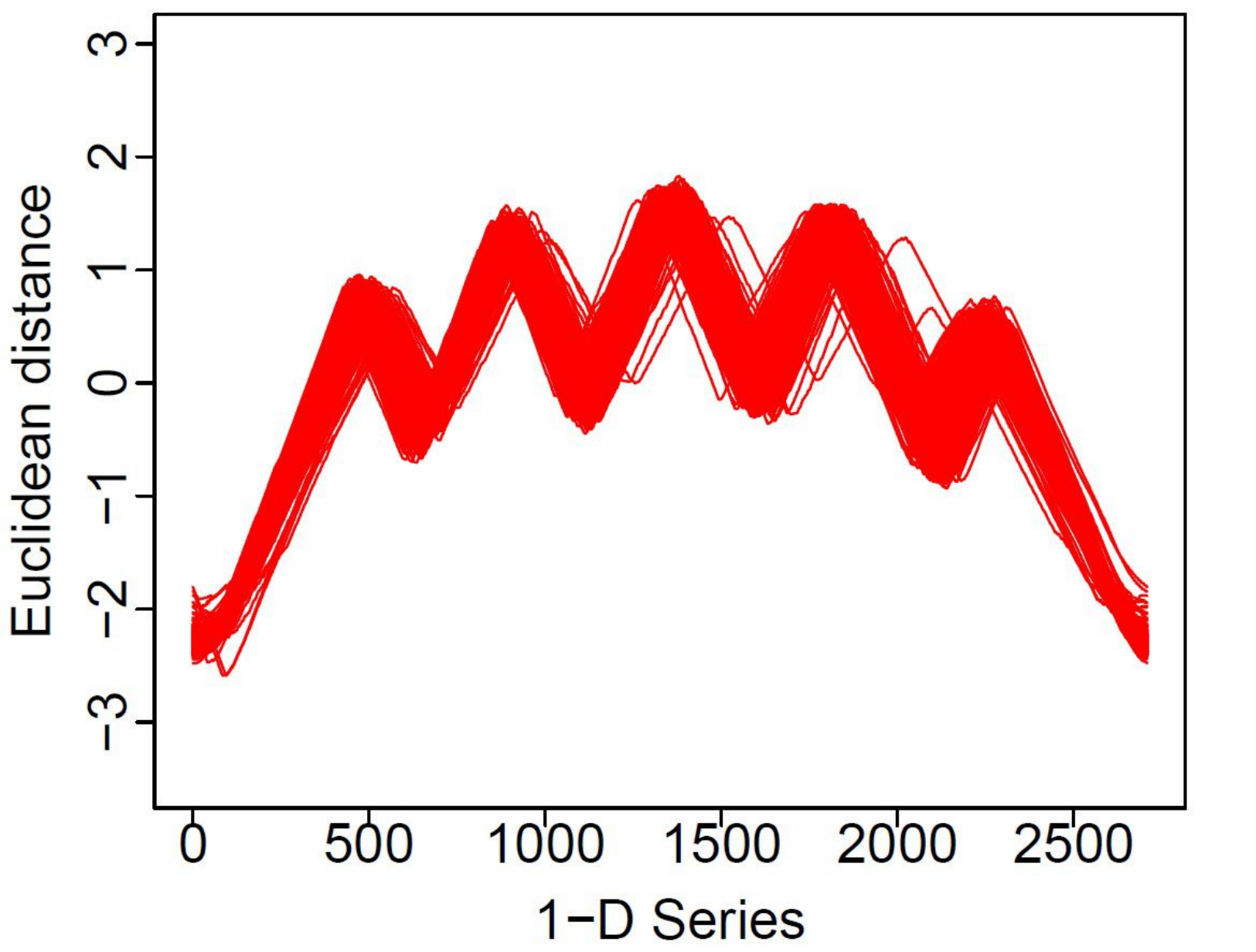}
\includegraphics[width=5.93cm]{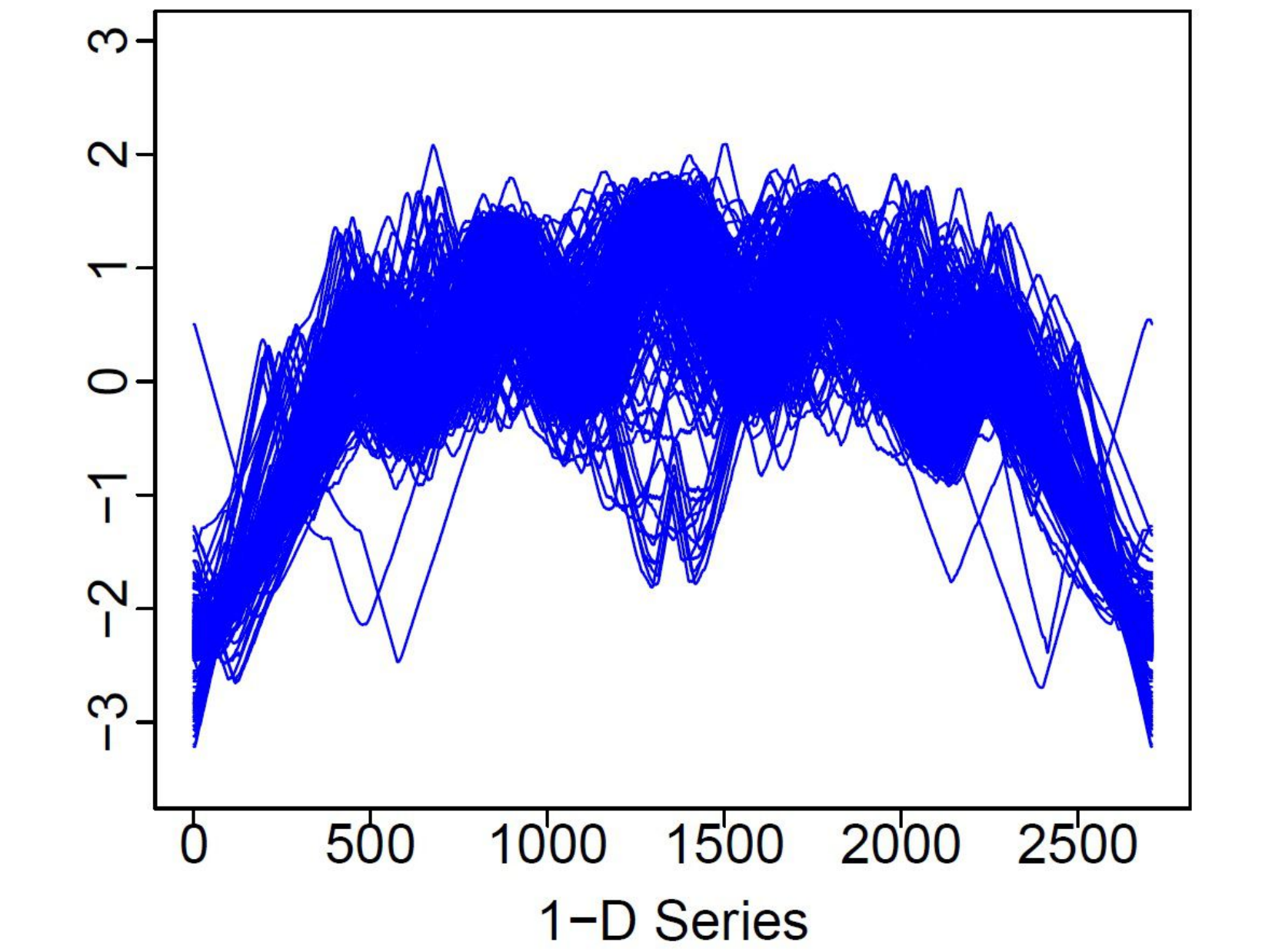}
\includegraphics[width=5.93cm]{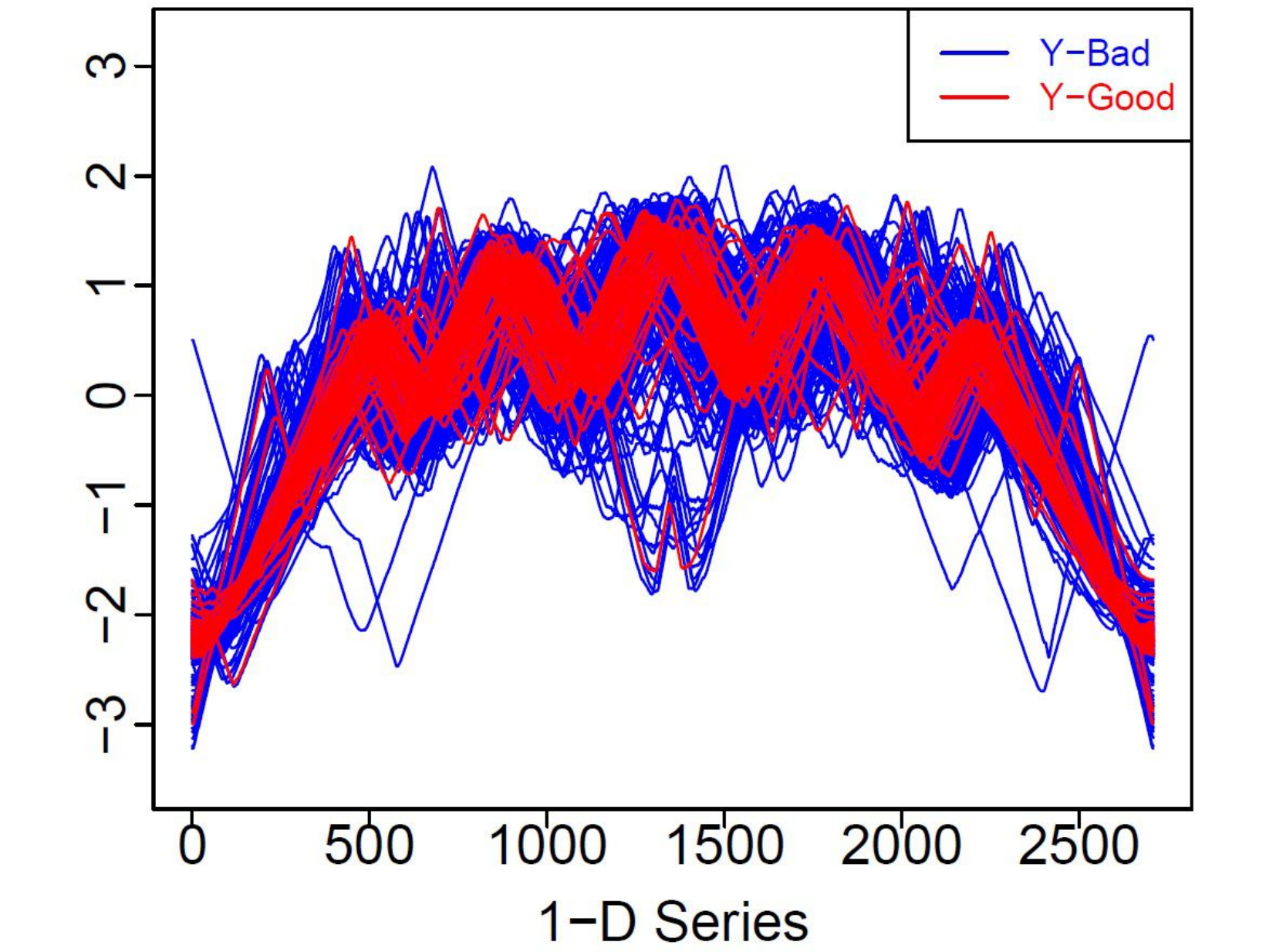}
\caption{\small{Euclidean distances for good hand outlines (left panel), bad hand outlines (middle panel), and Euclidean distances of hand outlines flagged as outlying (right panel)}.}\label{fig:Fig_4}
\end{figure}

We consider hand radiograph data, because this study is a part of a wider project aimed at predicting bone age using hand radiograph images, initially described by \cite{davis2012segmentation}. The dataset is available at \url{https://www.timeseriesclassification.com}. Determining bone age involves estimating the expected age of a patient from an X-ray image by measuring the development of non-dominant hand bones \citep{davis2012segmentation}. This approach evaluates how much a child's or individual's bones have developed at a certain time. The dataset, consisting of images, was created by digitizing X-ray films without altering the structure of the radiographs. A hand model was created based on the results obtained from the 1-D series using dynamic time warping. The radiographs were digitized using an Euclidean distance matrix to create the 1-D series. However, good hand outlines are required to determine bone ages correctly. With this dataset, we aim to classify whether a hand outline is good or bad. A graphical display of good and bad hand outlines and their Euclidean distances is presented in Figures~\ref{fig:Fig_3} and~\ref{fig:Fig_4}, respectively.

The dataset consists of 1370 observations, 495 bad hand outlines, and 875 good hand outlines. Euclidean distance measures are computed at 2709 equally spaced points in $[0, 2709]$. Several functional outlier detection methods, such as functional bagplot \citep{hyndman2010}, functional boxplot \cite{sun2011functional}, functional outlier map \citep{Rousseeuw2018}, and functional depth \citep{Febrero2008, Pintado2014, Nagy2017} can be used to determine the functional outliers in the hand radiograph data. In this study, we used the functional boxplot method of \cite{sun2011functional} to determine outliers in the dataset because of its simplicity. The results demonstrate that there are a total of 309 (22.5\%) outlying observations, of which 63 observations belong to the good hand outlines and 246 observations belong to the bad hand outlines (see Figure~\ref{fig:Fig_4}). This motivates us to apply our proposed method to make robust classification of hand outlines. Compared with non-robust methods, the proposed method is expected to provide improved classification performance for this dataset. In contrast, the proposed method is expected to provide similar or even better classification performance with less computing time than RFPLS. Note that the curves identified as outliers by the functional boxplot (or any other appropriate methods) may not necessarily be outliers. However, in this context, we consider these curves as potential outliers.

In our analyses, 200 Monte Carlo repetitions are performed. In each run, the dataset is divided into a training sample of randomly selected 959 observations (70\%) and a test sample of the remaining 30\% of the data. Models are fitted on the training sample, and the values of the binary response variable in the test sample are predicted to compute the AUC values. The median AUC values, along with standard errors given in brackets, are 0.770 (0.021), 0.794 (0.020), 0.788 (0.022), and 0.807 (0.017) for the FPLS, FPCA, RFPLS, and RFPCA methods, respectively. The results demonstrate that the proposed RFPCA method produces improved classification performance over the existing methods. The FPCA produces competitive AUC results to the proposed method. The worst classification results are produced by the FPLS method. The poor classification results produced by the FPLS and RFPLS methods may be because of the number of latent components determined by the iterative algorithms. 

As an alternative approach, we remove the identified outliers from the dataset and perform 200 Monte Carlo repetitions as described above to compute the AUC values. Our results show that the median AUC values, along with standard errors given in brackets, are 0.816 (0.024), 0.803 (0.025), 0.812 (0.023), and 0.811 (0.026) for the FPLS, FPCA, RFPLS, and RFPCA methods, respectively. From the results, it is observed that non-robust methods, that is FPLS and FPCA, yield superior AUC values compared to robust methods. The highest classification accuracy is achieved by the FPLS method. The proposed method demonstrates competitive classification results with both FPLS and FPCA. Upon removing outliers from the dataset, the classification accuracy of all methods increases, particularly for the non-robust methods. In comparison with FPLS and FPCA, the classification accuracy of RFPLS and RFPCA shows no significant change. This outcome underscores the crucial role of robust estimation when dealing with datasets contaminated by outliers.

\begin{figure}[!htb]
\centering
\includegraphics[width=4.42cm]{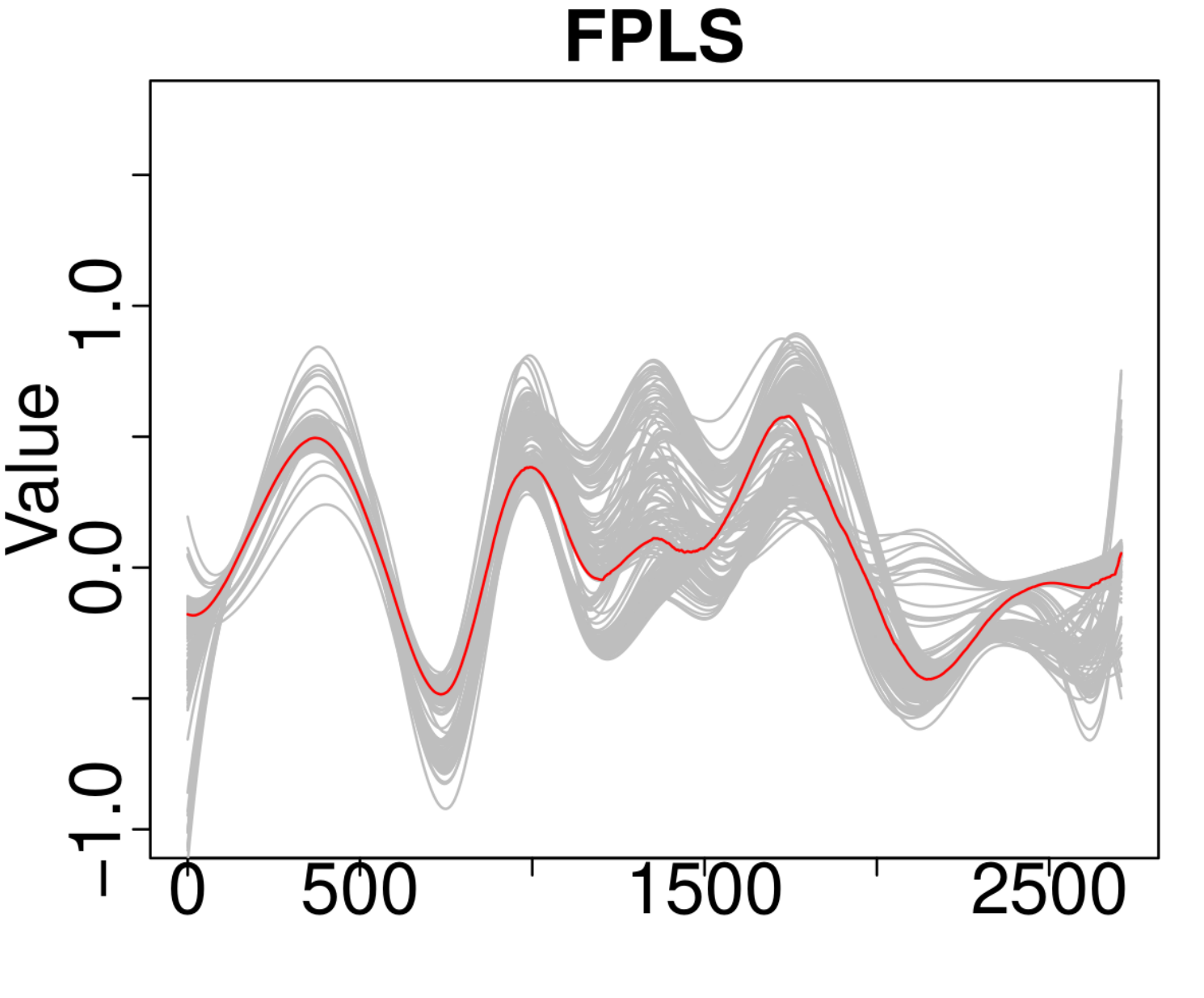}
\includegraphics[width=4.42cm]{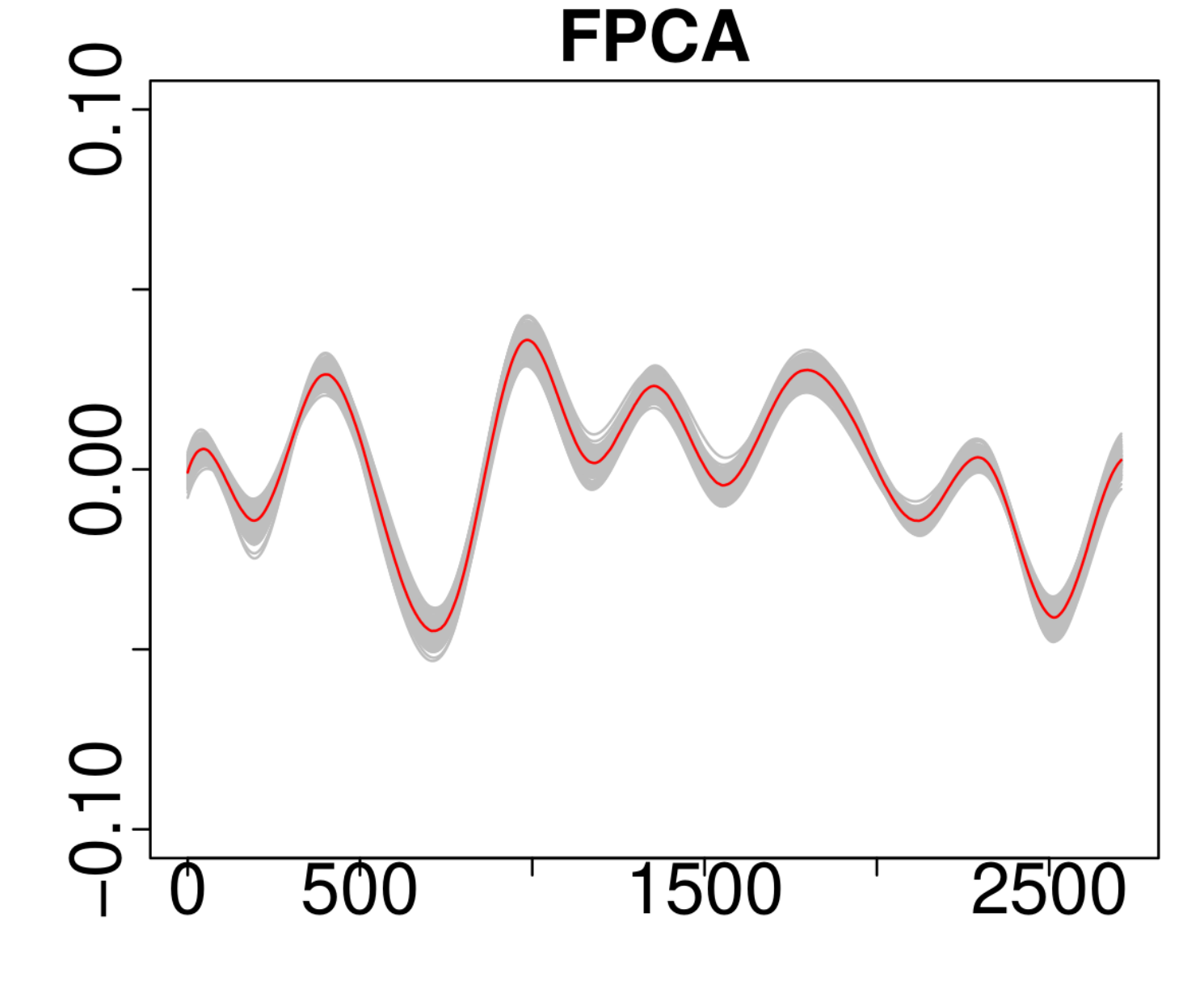}
\includegraphics[width=4.42cm]{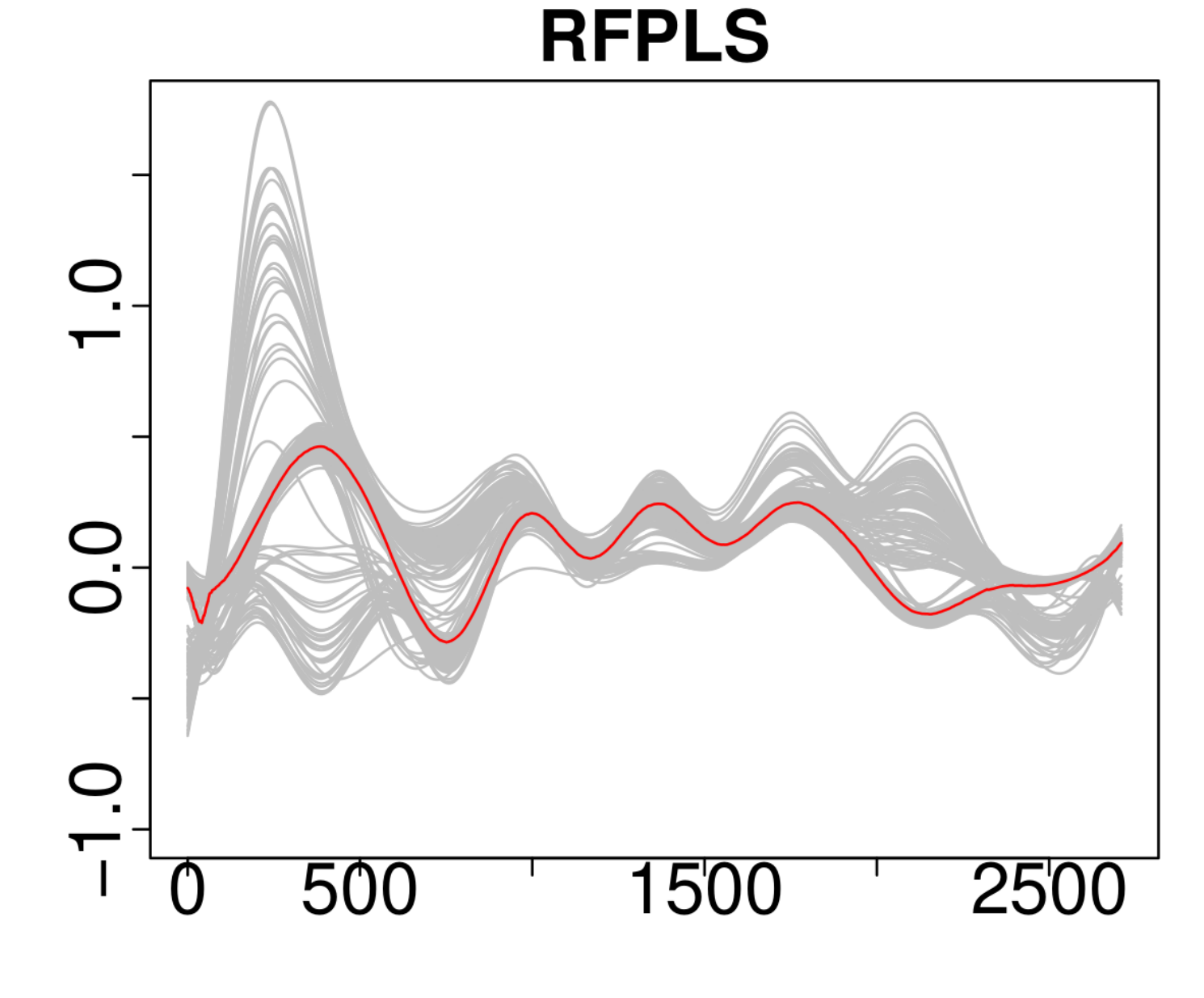}
\includegraphics[width=4.42cm]{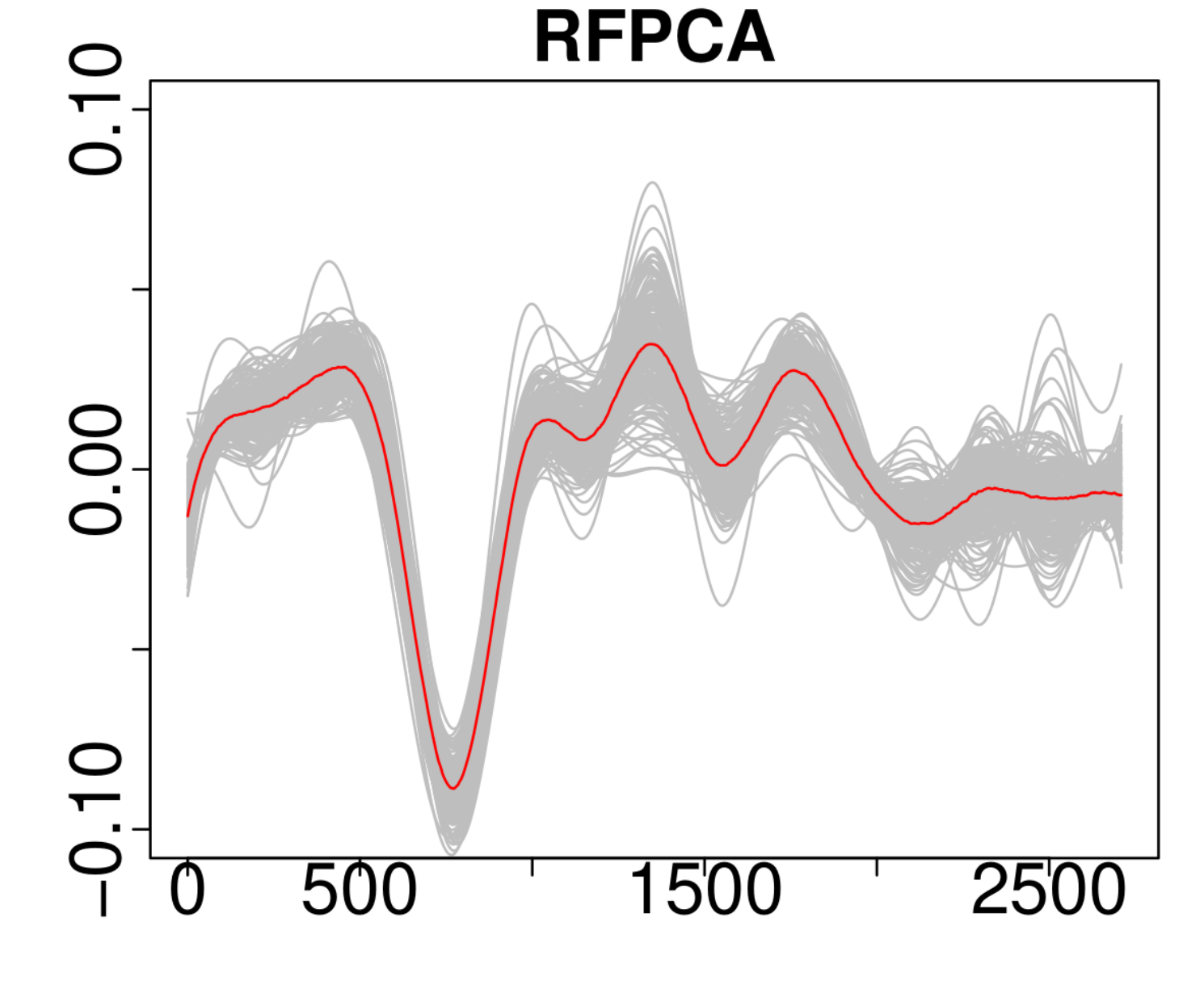} \\
\includegraphics[width=4.42cm]{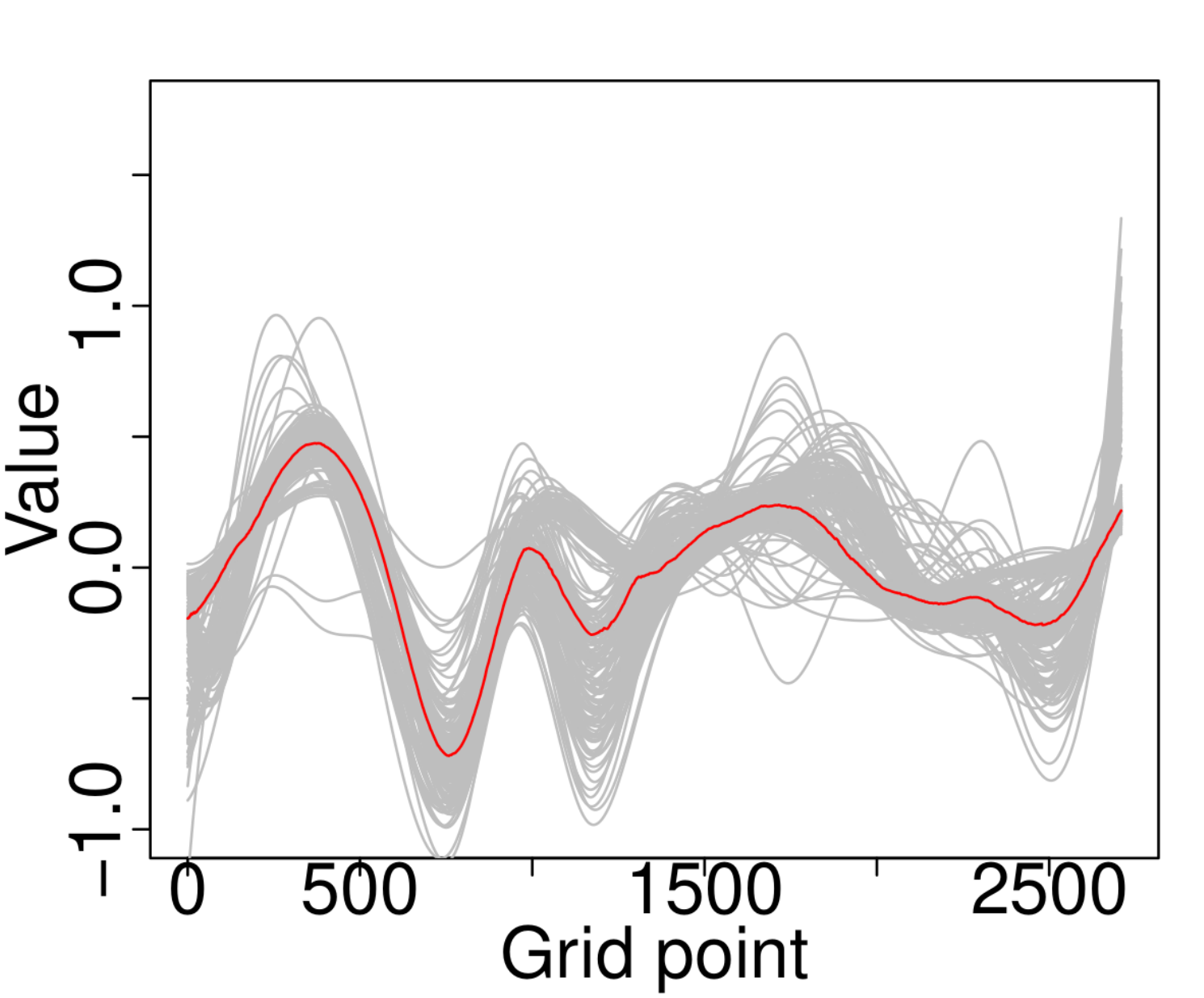}
\includegraphics[width=4.42cm]{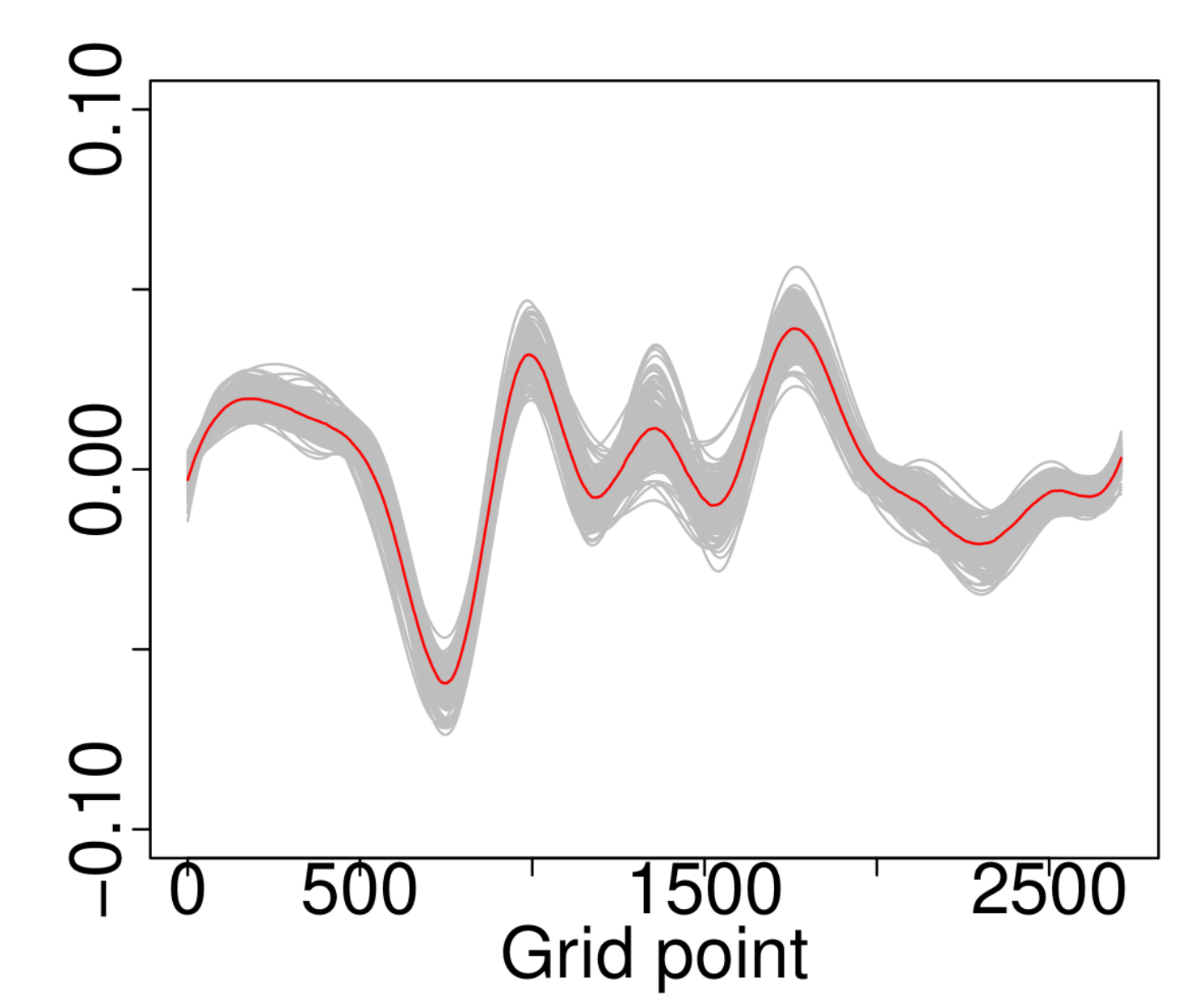}
\includegraphics[width=4.42cm]{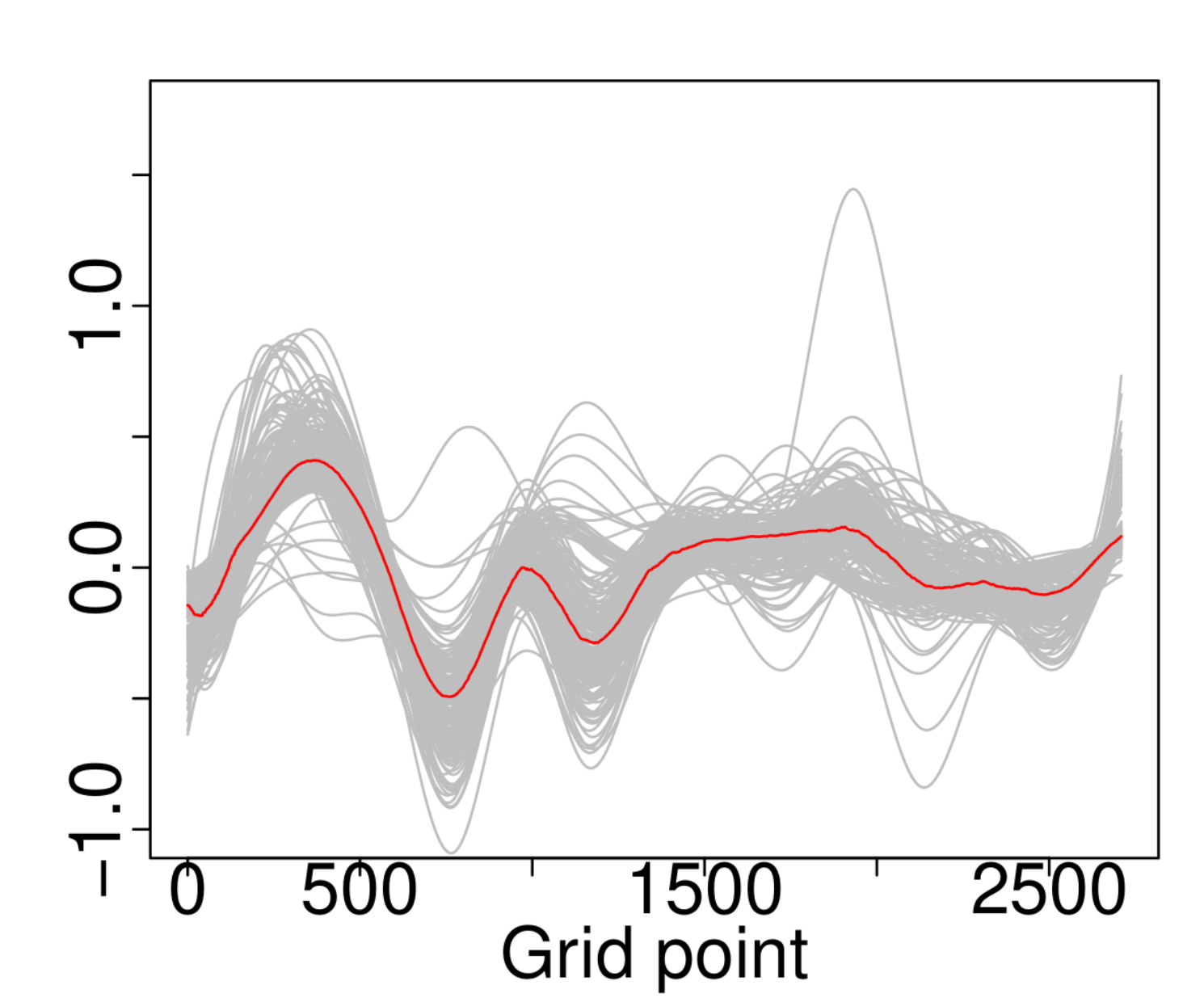}
\includegraphics[width=4.42cm]{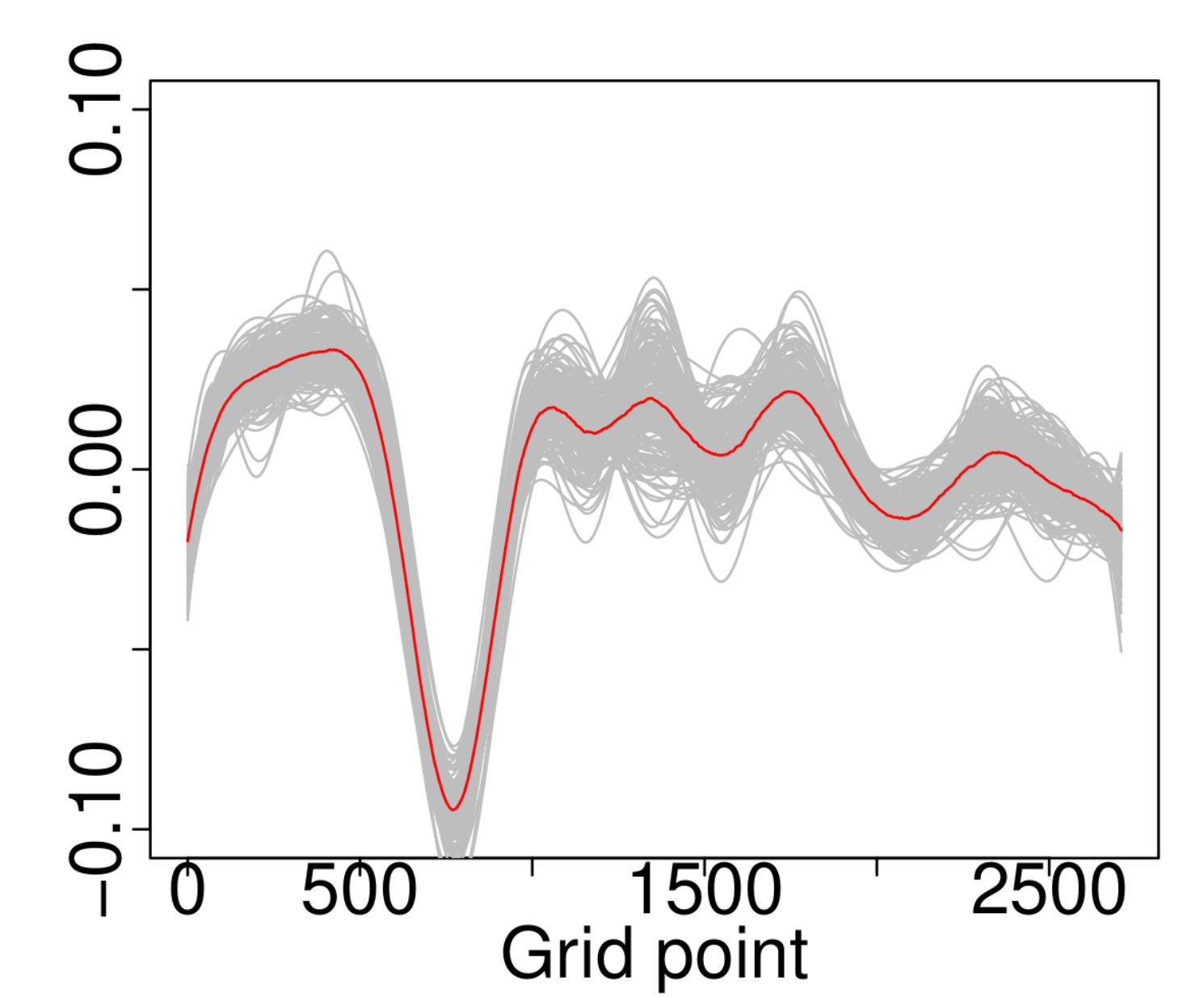} 
\caption{\small{Estimated regression coefficient functions (grey curves) and their averages (red curves) for hand radiograph data. The first row displays results before the removal of outliers, while the second row illustrates outcomes after the removal of outliers. Parameter functions are obtained over 200 Monte-Carlo runs for the FPLS (first panels), FPCA (second panels), RFPLS (third panels), and RFPCA (fourth panels) methods.}}\label{fig:Fig_5}
\end{figure}

The plots of the estimated regression coefficient function (before and after the removal of outliers) over 200 Monte-Carlo runs are presented in Figure~\ref{fig:Fig_5}. It is observed that FPCA-based methods produce similar and more stable parameter functions compared to FPLS-based methods. In Figure~\ref{fig:Fig_5}, the estimated regression coefficient functions for FPLS, FPCA, and RFPLS exhibit noticeable changes before and after the removal of outliers. In contrast, the changes in the estimated regression coefficient functions for the proposed RFPCA method are of a lesser degree.  Note that the $y$-axes of the panels in Figure~\ref{fig:Fig_5} are different for FPLS and FPCA-based methods, and the FPLS-based methods produce unstable parameter estimates (because of the number of latent components determined by the iterative algorithms).  Furthermore, the computing times required for the FPLS, FPCA, RFPLS, and RFPCA methods are recorded from a single Monte-Carlo run, and the computing times for these methods are obtained as 0.20, 0.24, 67.03, and 0.91 seconds, respectively. 

\section{Conclusion}\label{sec:conc}

Various methods based on pre-determined and data-driven bases have been proposed to estimate the parameters of FlogitR. While most existing methods perform well in cases where the data follow a smooth process, they may produce biased parameter estimates and decreased classification accuracy in the presence of outliers. Among the existing robust methods, the RFPLS has good classification performance in the presence of outliers but could be more computationally efficient. In this study, we propose the RFPCA method, which is both computationally efficient and has improved classification accuracy in the presence of outliers.

The estimation and classification performance of the proposed method is evaluated through a series of Monte-Carlo experiments and empirical data analysis. The results obtained by the proposed method are compared with those of existing non-robust FPLS and FPCA methods and the robust RFPLS method. We find that the proposed method exhibits similar or even better estimation and classification performance to the FPLS and FPCA methods when outliers are present in the dataset. Our results also indicate that the proposed method produces similar or superior results compared to the RFPLS method with significantly less computing time. The asymptotic consistency and influence function of the proposed robust method are investigated under mild conditions.

The proposed method can be further extended in at least three ways. First, in the present study, we consider a single functional predictor variable in FlogitR. Our proposed method can be extended to further improve estimation and classification accuracies where multiple functional predictors are used to estimate FlogitR. Second, we consider only functional predictor variables in the model. However, the binary response variable may depend on scalar predictor variables. Hence, our proposed method can be extended to a partial FlogitR that includes functional and scalar predictor variables. Third, the classification accuracy of the proposed method can be improved by utilizing a penalization term in the model to control the smoothness of the functional variables; that is, a penalized version of our proposed method can be proposed to obtain improved estimation and classification results for FlogitR.

\section*{Acknowledgments}

We thank two reviewers for their valuable comments and suggestions, which have improved our manuscript.

\newpage
\begin{center}
\large Appendixes
\end{center}
\appendix
We showcase the consistency and influence function of the estimator introduced in~\eqref{eq:propest}. Furthermore, we provide insights into deriving the asymptotic distribution of the aforementioned estimator. To establish the consistency of our proposed estimator, we outline the necessary conditions.

\begin{itemize}
\item[$C_1$.] $\X(t)$ has finite-dimensional Karhunen Lo\`{e}ve decompositions, i.e., $\X(t) = \sum_{k=1}^K \xi_k \psi_k(t)$ with eigenvalues $\lambda_1 > \ldots > \lambda_K > 0$.
\item[$C_2$.] The regression coefficient function $\beta(t)$ lies in a linear subspace spanned by $\{\psi_1(t), \ldots, \psi_K(t) \} \in \mathcal{L}_2(\mathcal{T})$.
\item[$C_3$.] The random variables $\{\xi_1, \ldots, \xi_K \}$ associated with $\{\psi_1(t), \ldots, \psi_K(t) \}$ are absolutely continuous with joint density $g(x)$ satisfying $g(x) = h(\Vert x \Vert_E)$ for $x \in \mathbb{R}^K$ for some measurable function $h: \mathbb{R} \rightarrow \mathbb{R}_{+}$, where $\Vert \cdot \Vert$ denotes the Euclidean norm.
\item[$C_4$.] $\X$ has finite fourth moments, i.e. $\Vert \X \Vert^4 < \infty$.
\item[$C_5$.] There is overlap in the sample, i.e., the overlap condition of \cite{albert1984} for the existence of the maximum likelihood estimator holds.
\item[$C_6$.] There exists $\eta > 0$ such that $\rho_2$ is increasing on $] - \infty, -\eta ]$ and either decreasing on $[\eta, + \infty [$.
\item[$C_7$.] $\lim_{\kappa \rightarrow \infty} \rho_2^{\prime}(\kappa;v) / \rho_2^{\prime}(- \kappa) = 0$ $\forall~ v > 0$.
\end{itemize}

Conditions $C_1$ and $C_2$ are essential for guaranteeing that a finite number of truncation constants, denoted as $K$, can effectively define the infinite-dimensional FLogitR. Meanwhile, Condition $C_3$ plays a crucial role in demonstrating the Fisher-consistency of the RFPCA estimator. The fulfilment of Conditions $C_1$ and $C_3$ is contingent upon the validity of $C_4$. To establish the Fisher-consistency of the WBY estimator, it is imperative to adhere to Conditions $C_5-C_7$.

\section{Consistency}

In the proposed FlogitR framework, the estimation of $\widehat{\beta}(t)$ is obtained through the utilization of the WBY-estimator within the RFPCA eigenspace, defined by the robust eigenfunctions of the covariance operator. Notably, these estimates may deviate significantly from those obtained using the maximum likelihood estimator within the FPCA eigenspace, which is spanned by the eigenfunctions, especially in the presence of outliers. Nonetheless, under specific regularity conditions governing the processes $\X(t)$, both estimates are consistent when outliers are absent in the data, as demonstrated in the ensuing lemma and remark. The ensuing lemma establishes the Fisher-consistency of the proposed estimator.

\begin{lemma}\label{lem1}
Under conditions $C_1-C_4$, the RFPCA estimator $\widehat{\beta}(t)$ is Fisher-consistent.
\end{lemma} 

\begin{proof}[Proof of Lemma~\ref{lem1}]
In the proof, we employ methodologies akin to those elucidated by \cite{croux2003implementing} and \cite{kalogridis2019}. Let $\mathbb{P}$ represent the image measure of $\X$, denoted by $\mathbb{P}(U) = P(\X \in U)$ for a Borel set $U$. Subsequently, the distribution function of $\X$ is defined as follows:
\begin{equation*}
\mathcal{F}(d_1, \ldots, d_K) := \mathcal{P}(\xi_1 \leq d_1, \ldots, \xi_K \leq d_K).
\end{equation*}
Then, the functional of the proposed robust estimator $\widehat{\beta}(t)$ is defined by:
\begin{equation*}
\widehat{\beta}(\mathcal{F}) = \sum_{k=1}^K \widehat{\gamma}_k(\mathcal{F}) \widehat{\psi}_k(\mathcal{F})(t).
\end{equation*}
Subsequently, the functional $\widehat{\beta}(t)$ is said to be Fisher-consistent under the condition that $\widehat{\beta}(\mathcal{F}) = \beta(t)$ for all $t \in \mathcal{T}$, indicating that $\widehat{\gamma}_k(\mathcal{F}) = \gamma_k$ and $\widehat{\psi}_k(\mathcal{F})(t) = \psi_k(t)$. In simpler terms, the functional $\widehat{\beta}(t)$ is considered Fisher-consistent if the M-estimator of location for functional data, RFPCA estimator, and WBY estimator all exhibit Fisher-consistency.

In accordance with \cite{kalogridis2019}, condition $C_3$ is sufficient for ensuring the Fisher-consistencies of both the M-estimator of location for functional data and the RFPCA estimator. Subsequently, under condition $C_1$, the expression $\X(t) = \sum_{k=1}^K \xi_k \widehat{\psi}_k(\mathcal{F})(t)$ holds. For the FlogitR, leveraging conditions $C_1$ and $C_2$ alongside the orthonormal properties of $\widehat{\psi}_k(\mathcal{F})(t)$, we obtain:
\begin{equation*}
l_i = \beta_0 + \sum_{k=1}^K \xi_k \widehat{\psi}_k(\mathcal{F})(t) \widehat{\psi}^\top_k(\mathcal{F})(t) \widehat{\gamma}_k(\mathcal{F}) = \beta_0 + \bm{\Xi}_i^\top \widehat{\bm{\gamma}}(\mathcal{F}), \quad i = 1, \ldots, n,
\end{equation*}
Let $\bm{Z} = [\bm{Z}_1, \ldots, \bm{Z}_n]^\top$, where $\bm{Z}_i = (1, \bm{\Xi}i^\top)^\top$, and $\bm{\theta} = [\beta_0, \widehat{\bm{\gamma}}(\mathcal{F})]$. According to Lemma 1 in \cite{croux2003implementing}, under the stipulations of Conditions $C_5-C_7$, the existence and finite norm of the Bianco and Yohai estimator are guaranteed. Let $H_0$ represent the model distribution, such that $P_{H_0}(Y = 1, \bm{Z} = \bm{z}) = F(\bm{Z}^\top \bm{\theta})$, where $F$ denotes the logistic function, i.e., $F(u) = 1 / (1 + \exp(-u))$. Under the model distribution $H_0$, following the Fisher-consistency of the Bianco and Yohai estimator with $\rho_2$ as outlined in \cite{croux2003implementing}, Conditions $C_5-C_7$ are sufficient to establish that $\text{E}_{H_0}[\rho_2^{\prime}(\bm{Z}^\top \bm{\theta}, Y \vert \bm{\Xi}^\top] = 0$. This substantiates the Fisher-consistency of the Bianco and Yohai estimator for $\bm{\theta}$. The WBY estimator, being a weighted variant of the Bianco and Yohai estimator, also maintains Fisher-consistency without necessitating additional assumptions, given that the weighting is solely determined by the values of $\bm{\Xi}$.

Taking into account the Fisher-consistencies of the M-estimator of location for functional data, RFPCA estimator, and WBY estimator collectively, the ultimate estimator $\widehat{\beta}(t)$ in Equation~\eqref{eq:propest} demonstrates Fisher-consistency, signifying that $\widehat{\beta}(\mathcal{F}) = \beta(t)$ for all $t \in \mathcal{T}$.
\end{proof}

\begin{remark}
Consider an i.i.d. random sample $\left\lbrace Y_i, \X_i(t): i = 1, \ldots, n \right\rbrace$ characterized by distribution functions $(\text{Be}(1, \pi_i), \mathcal{F})$. Under the assumption of condition $C_1$, the definitions of $\X_i(t)$ involve finite numbers of eigenfunctions and their corresponding random variables. The Glivenko-Cantelli theorem \citep[refer, for instance, to][]{polard1984} facilitates the demonstration that the empirical distribution functions of these random variables uniformly converge to their population distribution functions. In the context of Lemma~\eqref{lem1} and the adherence to conditions $C_1-C_7$, it can be asserted that $\widehat{\beta}_(t)$ achieves asymptotic consistency. This assertion stems from the fact that Fisher-consistency is synonymous with asymptotic consistency, given the convergence of empirical distribution functions of random variables to their population distribution functions in a point-wise manner.
\end{remark}

\section{Influence function}

The evaluation of our proposed estimator's resistance to outliers will be conducted using the influence function approach introduced by \cite{Hampel}. This method quantifies the rate of asymptotic bias of an estimator in response to infinitesimal contamination in the distribution. A bounded influence function indicates robustness against extreme outliers. Let $(Y, \X^\top)^\top$ be a random variable with distribution $F_\theta=(G_0 \times \mathbb{P})$, where $\bm{\theta} \in \bm{\Theta}$. Consider an estimating functional $T$ of $\bm{\theta}$ that is Fisher-consistent, i.e., $T(F_{\bm{\theta}}) = \bm{\theta}$. The influence function of $T$ is then defined as:
\begin{equation}
	IF(\bm{v}, T, F_{\bm{\theta}}) = \lim_{\epsilon \downarrow 0}\frac{T(F_{{\bm{\theta}}, \epsilon}) -T(F_{\bm{\theta}}) }{\epsilon} = \frac{\partial}{\partial \epsilon} T(F_{\bm{\theta}})\Big|_{\epsilon=0} ,
\end{equation}
where $F_{\bm{\theta}, \epsilon} = (1-\epsilon) F_{\bm{\theta}} + \epsilon \delta_{\bm{v}}$ is the contaminated distribution with $\delta_{\bm{v}}$ being the point-mass distribution at  $\bm{v}=(Y_0, \X_0)$. Thus, the influence function is the Gateaux derivative of the functional $T$ defined on the space of finite signed measures in the direction $\delta_{\bm{v}} - F_{\bm{\theta}}$.

Let us consider the WBY estimator:
\begin{equation*}
\widehat{\bm{\theta}} = (\widehat{\beta}_0, \widehat{\gamma}^\top)^\top= \arg \min_{\bm{\theta}} \sum_{i=1}^n \omega_i \left\{
 \rho_2[d(\bm{Z}_i^\top \bm{\theta}; Y_i)] + C(\bm{Z}_i^\top \bm{\theta})\right\}.
\end{equation*}
The estimating equation for the WBY estimator is then given by
\begin{equation*}
 \sum_{i=1}^n \omega_i \left\{
 \rho_2^{\prime}[d(\bm{Z}_i^\top \bm{\theta}; Y_i)] d^{\prime}(\bm{Z}_i^\top \bm{\theta}; Y_i) + C^{\prime}(\bm{Z}_i^\top \bm{\theta})\right\} \bm{Z}_i = 0.
\end{equation*}
The population version of the estimating equation is as follows:
\begin{equation}
 \int_y \int_x \omega \left\{
 \rho_2^{\prime}[d(\bm{z}^\top \bm{\theta}; y)] d^{\prime}(\bm{z}^\top \bm{\theta}; y) + C^{\prime}(\bm{z}^\top \bm{\theta})\right\} \bm{z} d F_{\bm{\theta}} = 0.
 \label{est_eqn}
\end{equation}

Let $\bm{\theta}_\epsilon$ be the value of the parameter under the contaminated distribution $F_{\bm{\theta}, \epsilon} = (1-\epsilon) F_{\bm{\theta}} + \epsilon \delta_{\bm{v}}$, where $\bm{v}=(Y_0, \X_0)$. Then, the above estimating equation can be written as
\begin{equation*}
 \int_y \int_x \omega \left\{
 \rho_2^{\prime}[d(\bm{z}^\top \bm{\theta}_\epsilon; y)] d^{\prime}(\bm{z}^\top \bm{\theta}_\epsilon; y) + C^{\prime}(\bm{z}^\top \bm{\theta}_\epsilon)\right\} \bm{z} d F_{\bm{\theta}_\epsilon} = 0.
\end{equation*}
Differentiating the above equation with respect to $\epsilon$, we get
\begin{align*}
 & \int_y \int_x \omega \left\{
 \rho_2^{\prime \prime}[d(\bm{z}^\top \bm{\theta}_\epsilon; y)] (d^{\prime})^2(\bm{z}^\top \bm{\theta}_\epsilon; y) + C^{\prime \prime}(\bm{z}^\top \bm{\theta}_\epsilon)\right\} \left\{ \bm{z}^\top \frac{\partial \bm{\theta}_\epsilon}{\partial \epsilon} + \bm{\theta}_\epsilon^\top \frac{\partial \bm{z}}{\partial \epsilon} \right\} \bm{z} d F_{\bm{\theta}_\epsilon} \\
& + \int_y \int_x \omega 
 \rho_2^{\prime}[d(\bm{z}^\top \bm{\theta}_\epsilon; y)] d^{\prime \prime}(\bm{z}^\top \bm{\theta}_\epsilon; y) \left\{ \bm{z}^\top \frac{\partial \bm{\theta}_\epsilon}{\partial \epsilon} + \bm{\theta}_\epsilon^\top \frac{\partial \bm{z}}{\partial \epsilon} \right\}  \bm{z} d F_{\bm{\theta}_\epsilon}\\
 & \int_y \int_x \omega \left\{
 \rho_2^{\prime}[d(\bm{z}^\top \bm{\theta}_\epsilon; y)] d^{\prime}(\bm{z}^\top \bm{\theta}_\epsilon; y) + C^{\prime}(\bm{z}^\top \bm{\theta}_\epsilon)\right\} \frac{\partial \bm{z}}{\partial \epsilon} d F_{\bm{\theta}_\epsilon} \\
& + \int_y \int_x \omega \left\{
 \rho_2^{\prime}[d(\bm{z}^\top \bm{\theta}_\epsilon; y)] d^{\prime}(\bm{z}^\top \bm{\theta}_\epsilon; y) + C^{\prime}(\bm{z}^\top \bm{\theta}_\epsilon)\right\} \bm{z} (-dF_{\bm{\theta}} + d\delta_{\bm{v}}) = 0.
\end{align*}
Therefore, we have
\begin{align*}
&\frac{\partial \bm{\theta}_\epsilon}{\partial \epsilon} = - \left[ 
\int_y \int_x \omega \left\{
 \rho_2^{\prime \prime}[d(\bm{z}^\top \bm{\theta}_\epsilon; y)] (d^{\prime})^2(\bm{z}^\top \bm{\theta}_\epsilon; y) + \rho_2^{\prime}[d(\bm{z}^\top \bm{\theta}_\epsilon; y)] d^{\prime \prime}(\bm{z}^\top \bm{\theta}_\epsilon; y) +  C^{\prime \prime}(\bm{z}^\top \bm{\theta}_\epsilon)\right\}  \bm{z} \bm{z}^\top d F_{\bm{\theta}_\epsilon}
\right]^{-1}\\
& \times \Bigg[
\int_y \int_x \omega \left\{
 \rho_2^{\prime \prime}[d(\bm{z}^\top \bm{\theta}_\epsilon; y)] (d^{\prime})^2(\bm{z}^\top \bm{\theta}_\epsilon; y) + \rho_2^{\prime}[d(\bm{z}^\top \bm{\theta}_\epsilon; y)] d^{\prime \prime}(\bm{z}^\top \bm{\theta}_\epsilon; y) +  C^{\prime \prime}(\bm{z}^\top \bm{\theta}_\epsilon)\right\} \bm{\theta}_\epsilon^\top \frac{\partial \bm{z}}{\partial \epsilon} \bm{z}  d F_{\bm{\theta}_\epsilon}\\
 & +  \int_y \int_x \omega \big\{
 \rho_2^{\prime}[d(\bm{z}^\top \bm{\theta}_\epsilon; y)] d^{\prime}(\bm{z}^\top \bm{\theta}_\epsilon; y) + C^{\prime}(\bm{z}^\top \bm{\theta}_\epsilon)\big\} \frac{\partial \bm{z}}{\partial \epsilon} d F_{\bm{\theta}_\epsilon}  + \omega \big\{
 \rho_2^{\prime}[d(\bm{z}^\top \bm{\theta}_\epsilon; y)] d^{\prime}(\bm{z}^\top \bm{\theta}_\epsilon; y) + C^{\prime}(\bm{z}^\top \bm{\theta}_\epsilon)\big\} \bm{z} \bigg|_{\bm{v}}
\Bigg].
\end{align*}
Note that the last term with integral $dF_{\bm{\theta}}$ vanishes from~\eqref{est_eqn}. Now, evaluating both side at $\epsilon=0$, we get the influence function:
\begin{align}
IF(\bm{v}, \theta, F_{\bm{\theta}}) &= - \left[ 
\int_y \int_x \omega \left\{
 \rho_2^{\prime \prime}[d(\bm{z}^\top \bm{\theta}; y)] (d^{\prime})^2(\bm{z}^\top \bm{\theta}; y) + \rho_2^{\prime}[d(\bm{z}^\top \bm{\theta}; y)] d^{\prime \prime}(\bm{z}^\top \bm{\theta}; y) +  C^{\prime \prime}(\bm{z}^\top \bm{\theta})\right\}  \bm{z} \bm{z}^\top d F_{\theta}
\right]^{-1} \nonumber\\
& \times \Bigg[
\int_y \int_x \omega \left\{
 \rho_2^{\prime \prime}[d(\bm{z}^\top \bm{\theta}; y)] (d^{\prime})^2(\bm{z}^\top \bm{\theta}; y) + \rho_2^{\prime}[d(\bm{z}^\top \bm{\theta}; y)] d^{\prime \prime}(\bm{z}^\top \bm{\theta}; y) +  C^{\prime \prime}(\bm{z}^\top \bm{\theta})\right\} \bm{\theta}^\top R \bm{z}  d F_{\bm{\theta}} \nonumber\\
 & +  \int_y \int_x \omega \left\{
 \rho_2^{\prime}[d(\bm{z}^\top \bm{\theta}; y)] d^{\prime}(\bm{z}^\top \bm{\theta} y) + C^{\prime}(\bm{z}^\top \bm{\theta})\right\} R d F_{\bm{\theta}} \label{if}\\
 & + \omega \left\{
 \rho_2^{\prime}[d(\bm{z}^\top \bm{\theta}; y)] d^{\prime}(\bm{z}^\top \bm{\theta}; y) + C^{\prime}(\bm{z}^\top \bm{\theta})\right\} \bm{z} \bigg|_{\bm{v}}
\Bigg], \nonumber
\end{align}
where $R$ is a vector with first term as 0, and the remaining terms as $IF(\X_0, \xi, \mathbb{P})$, where $\xi = (\xi_1, \xi_2, \cdots, \xi_K)^T$. 

From Theorem 3.1 of \cite{bali2011rfpc}, we find the influence function of the $k^\textsuperscript{th}$ eigenvalue as:
\begin{equation*}
IF(\X_0, \xi_k, \mathbb{P}) =  2 \xi_k IF\left(\frac{\langle x, \psi_k \rangle}{\sqrt{\xi_k}}  ; \sigma_M, \mathbb{P} \right),
\end{equation*}
where $IF\left(\cdot  ; \sigma_M, \mathbb{P} \right)$ is the influence function of the M-scale function for $\sigma_M$:
\begin{equation*}
IF\left(u  ; \sigma_M, \mathbb{P} \right) = \left[ \rho_{1,c}(u) - \delta\right]  \left\{E_{\mathbb{P}}[\rho_{1,c}^{\prime}(\X)\X]\right\}^{-1}. 
\end{equation*}
As $\bm{\theta} = (\beta_0, \gamma^\top)^\top$, we get the influence function of the estimator of $\beta_0$ as the first component of $IF(\bm{v}, \bm{\theta}, F_{\bm{\theta}})$ from~\eqref{if}. The remaining component of~\eqref{if} gives $IF(\bm{v}, \gamma, F_{\bm{\theta}})$. Finally, we find the influence function of $\widehat{\beta}$ using the relation $\widehat{\beta} = \Xi^\top \widehat{\gamma}$ as
\begin{equation}
IF(\bm{v}, \beta, F_{\bm{\theta}}) = \Xi^\top IF(\bm{v}, \gamma, F_{\bm{\theta}}) + \gamma^\top IF(\X_0, \Xi, \mathbb{P}),
\end{equation}
where $IF(\X_0, \Xi, \mathbb{P})$ is the influence function of the eigenfunctions. Let $IF(\X_{k0}, \bm{\psi}_k, \mathbb{P})$ be the $k^\textsuperscript{th}$ component of $IF(\X_0, \Xi, \mathbb{P})$ for $k = 1, \ldots, K$. Then,  $IF(\X_{k0}, \bm{\psi}_k, \mathbb{P})$ is calculated using Theorem 3.1 of \cite{bali2011rfpc}. If  $\lambda_1 > \lambda_2 > \cdots > \lambda_q > \lambda_{q+1}$, then $\phi_j(t)$ are unique up to a sign change when $1\leq j\leq q$. Then, the influence function of the $k^\textsuperscript{th}$ eigenfunction $\bm{\phi}_k$ for $k\leq q$,  is given by
\begin{align*}
	IF(\X_{0}, \bm{\phi}_k, \mathbb{P}) &= \sum_{j \geq k +1} \frac{\sqrt{\lambda_k}}{\lambda_k - \lambda_j} DIF\left( \frac{ \langle \X, \phi_k \rangle}{\sqrt{\lambda_k}}, \sigma_M, \mathbb{P}\right) \langle \X, \phi_j\rangle \phi_j \\
	& + \sum_{j = 1}^{k-1} \frac{\sqrt{\lambda_j}}{\lambda_k - \lambda_j} DIF\left( \frac{ \langle \X, \phi_j\rangle}{\sqrt{\lambda_j}}, \sigma_M, \mathbb{P} \right) \langle \X, \phi_k\rangle \phi_j, \nonumber
\end{align*}
where the derivative of the influence function of $\widehat{\sigma}_M$ is given by
\begin{equation*}
	DIF\left( u, \sigma_M, \mathbb{P} \right) =  \rho_{1,c}^{\prime}(u) \left\{\text{E}_{\mathbb{P}}[\rho_{1,c}^{\prime}(\X)\X] \right\}^{-1},
\end{equation*}
and the function $\rho_{1,c}$ is given in~\eqref{eq:loss1}.

Note that $\rho_{1,c}^{\prime}(u)=0$ for $\vert u \vert > c$. So, $IF(\X_{0}, \bm{\phi}_k, \mathbb{P})$ tends to zero as $\Vert \X \Vert \rightarrow \infty$, indicating a redescending effect for large leverage points. However, this influence function could still be unbounded but only for small scores on some eigenfunctions and simultaneously large scores on others. In other words, as \cite{bali2011rfpc} mentioned, observations with large absolute values of $\langle \X, \phi_j\rangle$ in combination with small absolute values of $\langle \X, \phi_k\rangle$ for $k < j$ may exert significant influence on the eigenfunctions. 

\begin{remark}
Other than a robustness measure, the influence function is also used to calculate the asymptotic distribution of an estimator. Under some mild conditions as given in \cite{Fernholz}, a Bahadur expansion yields
\begin{equation*}
	\sqrt{n}(\widehat{\beta} -  \beta) = \frac{1}{\sqrt{n}} \sum_{i=1}^{n} 	IF((Y_i, \X_i), \beta, F_{\bm{\theta}}) + o_p(1).
\end{equation*}  
It gives the asymptotic distribution of our FlogitR estimator as:
\begin{equation*}
	\sqrt{n}(\widehat{\beta}-  \beta) \overset{a}{\sim} N(0, \bm{\Sigma}),
\end{equation*} 
where
\begin{equation*}
	\bm{\Sigma} = \text{E}_{F_{\bm{\theta}}} \left(IF((Y_1, \X_1), \beta, F_{\bm{\theta}}) \otimes	IF((Y_1, \X_1), \beta, F_{\bm{\theta}})\right).
\end{equation*}
Similarly, we can derive the asymptotic distribution of the intercept parameter $\widehat{ \beta}_0$ from its influence function.
\end{remark}

\newpage
\bibliographystyle{agsm}
\bibliography{rfpca.bib}

\end{document}